\documentclass[10pt]{iopart}
\usepackage{latexsym,epsfig,graphicx,bbm}

\newcommand{\half}{\mbox{$\textstyle \frac{1}{2}$}}
\newcommand{\ket}[1]{\left | \, #1 \right \rangle}
\newcommand{\bra}[1]{\left \langle #1 \, \right |}

\newcommand{\bracket}[3]{\left\langle #1 \left| #2 \right| #3 \right\rangle}

\newcommand{\av}[1]{\langle #1\rangle}
\newcommand{\outprod}[2]{\ket{#1}\bra{#2}}

\newcommand{\vac}{\ket{\textrm{vac}}}

\newcommand{\eqr}[1]{equation~(\ref{#1})}
\newcommand{\fir}[1]{figure~\ref{#1}}
\def \i {\textrm{i}}

\begin{document}

\paper[Graph state generation with noisy mirror-inverting spin
chains]{Graph state generation with noisy mirror-inverting spin
chains}

\author{Stephen~R~Clark, Alexander~Klein, Martin~Bruderer and Dieter~Jaksch}
\address{Clarendon Laboratory, University of Oxford, Oxford OX1 3PU, U.K.}

\ead{s.clark@physics.ox.ac.uk}

\begin{abstract}
We investigate the influence of noise on a graph state generation
scheme which exploits a mirror inverting spin chain. Within this
scheme the spin chain is used repeatedly as an entanglement bus
(EB) to create multi-partite entanglement. The noise model we
consider comprises of each spin of this EB being exposed to
independent local noise which degrades the capabilities of the EB.
Here we concentrate on quantifying its performance as a
single-qubit channel and as a mediator of a two-qubit entangling
gate, since these are basic operations necessary for graph state
generation using the EB. In particular, for the single-qubit case
we numerically calculate the average channel fidelity and whether
the channel becomes entanglement breaking, i.e., expunges any
entanglement the transferred qubit may have with other external
qubits. We find that neither local decay nor dephasing noise cause
entanglement breaking. This is in contrast to local thermal and
depolarizing noise where we determine a critical length and
critical noise coupling, respectively, at which entanglement
breaking occurs. The critical noise coupling for local
depolarizing noise is found to exhibit a power-law dependence on
the chain length. For two qubits we similarly compute the average
gate fidelity and whether the ability for this gate to create
entanglement is maintained. The concatenation of these noisy gates
for the construction of a five qubit linear cluster state and a
Greenberger-Horne-Zeilinger state indicates that the level of
noise that can be tolerated for graph state generation is tightly
constrained.
\end{abstract}



\maketitle

\section{Introduction}
Entanglement appears to be a crucial ingredient for the
potentially remarkable speedup of a quantum computer compared to
that of a classical
computer~\cite{Jozsa-PRSA-2003,Vidal-PRL-2003}. This observation
is especially highlighted within the one-way quantum computing
model~\cite{Raussendorf-PRL-2001,Raussendorf-PRA-2003}. Here the
state of a quantum many-body system, typically composed of
spin-$\half$ or qubit subsystems, can serve a universal resource
for quantum computing in which the computation is driven by
successive von-Neumann measurements on the individual
constituents. While the characterization of multipartite
entanglement in a general quantum many-body state remains an open
problem, initial states which can act as a universal resource for
one-way quantum computing are within an increasingly well-studied
class called {\em graph states}~\cite{Hein-PRA-2004,Hein-2006}.

Graph states are many-body quantum states which have an intuitive
representation in terms of mathematical graphs. More precisely,
vertices of a graph are assigned to the constituent qubits, each
initialized in a state $\ket{+} =(\ket{0} + \ket{1})/\sqrt{2}$,
and edges connecting vertices represent a pattern of Ising-type
interactions\footnote{This Ising interaction is typically taken to
implement a {\em controlled phase} or c-$\sigma^z$ gate.} that
have subsequently taken place between these qubits. In this way
the graph describes a preparation procedure for this class of
states, as depicted in \fir{spincomb}(a). Within the graph
formalism many of the properties of graph states, such as their
Schmidt measure and robustness to noise, can be computed
efficiently despite being intractable for a general
state~\cite{Hein-PRA-2004,Hein-PRA-2005,Hein-2006}. From such
studies it is known that there are graph states that contain the
maximum amount of entanglement permissible for any given number of
qubits. As such, graph states form a highly non-trivial class of
quantum states.

It is believed that some of the tremendous challenges faced in
realizing a quantum computer can be lessened by using an
architecture based on graph states~\cite{Hein-2006}. In particular
the underlying resource for one-way quantum computing is a special
class of graph states, called cluster
states~\cite{Briegel-PRL-2001}, which are represented by graphs
with a regular lattice geometry like that shown in
\fir{spincomb}(b). This pattern of nearest-neighbour Ising
interactions is a geometry which is very naturally suited to
quantum lattice systems. Additionally, by separating the
preparation of entanglement from its consumption within a
computation the one-way model can be arranged to accommodate lossy
or even probabilistic processes during the preparation phase.
Beyond cluster states more general graph states are also an
efficient resource for {\em specific} quantum
computations~\cite{Raussendorf-PRA-2003} and so represent a
preferred experimental route to quantum information processing
where qubits are a precious quantity. Graph states also play a
prominent role as code-words in quantum error
correction~\cite{Gottesman-1997} which permit the reliable storage
of quantum information in the presence of noise.

There are now a diverse range of proposals for the preparation of
graph states in realistic physical systems~\cite{Hein-2006}. These
include the direct use of linear optics and photon resolving
measurements to construct graph states with photons via a
non-deterministic protocol~\cite{Browne-PRL-2005}. As a proof of
principle an entirely optical creation of a 4 qubit graph state
was recently realized and used to implement a 2 qubit Grover
search algorithm~\cite{Walther-NAT-2005}. Other frameworks include
using hybrid systems which combine optical and solid state
qubits~\cite{Barrett-PRA-2005}. Another method is to instead
engineer a many-body quantum system whose ground state is a graph
state so that beyond engineering the nearest-neighbour
interactions the preparation becomes a cooling
problem~\cite{Bartlett-PRA-2006}. The approach which we consider
in this paper is based on exploiting a spin chain with fixed
engineered couplings chosen such that its dynamical evolution is
mirror
inverting~\cite{Christandl-PRL-2004,Christandl-PRA-2005,Cook-PRA-1979}.
Such spin chains have attracted much attention because of their
ability to perform perfect state transfer and therefore act as a
quantum communication
channel~\cite{Yung-PRA-2005,Karbach-PRA-2005,Yung-PRA-2006,Kay-PRL-2007}.
In reference~\cite{Clark-NJP-2005} it was shown that mirror
inverting spin chains are capable of implementing a specific type
of multi-qubit circuit that is naturally suited to the generation
of entanglement of the type present in graph states. For this
reason we call this type of chain an {\em entangling bus} (EB).
When the EB is used within a spin-ladder arrangement, as shown in
\fir{spincomb}(c) where the second leg of the ladder is a register
R of qubits, it permits the efficient generation of arbitrary
graph states within this register.

Experimental realizations of quantum systems inevitably possess a
coupling to a surrounding environment composed of a large number
of degrees of freedom which are beyond the experimenters
control~\cite{Breuer-2002}. This coupling introduces quantum noise
that destroys quantum coherence of the system (i.e. decoherence).
This is broadly classified as dissipation, when accompanied by the
exchange of energy between the system and environment, or
dephasing when there is no energy exchange. The effects of noise
on a spin chain used as quantum channels has been investigated
previously~\cite{DeChiara-PRA-2005,Cai-PRA-2006,Burgarth-PRA-2006,Zhou-2006}.
Here we consider a broader set of properties including the ability
of mirror inverting chains to both distribute and generate
entanglement which are crucial for the more challenging use of
them as EB. To do this we consider a specific, but physically
relevant~\cite{Briegel-PRA-1993,Hein-PRA-2005}, noise model where
each spin in the EB is weakly coupled to an independent
environment $\mathcal{E}$ and the complete chain is described by a
master equation, as illustrated in \fir{spincomb}(d).

\begin{figure}
\begin{center}
\includegraphics[width=9cm]{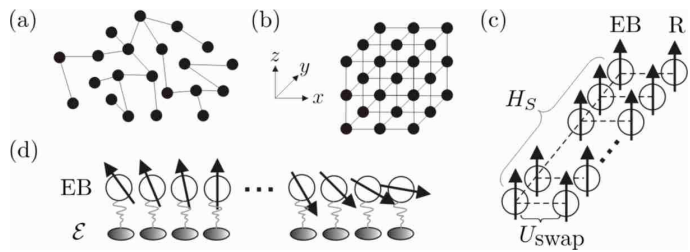}
\caption{(a) An arbitrary graph state. (b) A 3D cluster state. (c)
The spin-ladder arrangement used in the graph state generation
scheme. One leg of the ladder is the EB spin chain with a mirror
inverting Hamiltonian $H_S$. The other is a chain of decoupled
spins which form a storage register R. Coupling between adjacent
spins in EB and R is dynamically controlled to implement a rapid
swap gate $U_{\mathrm{swap}}$. (d) The EB spin chain with each
spin exposed to an independent local environment
$\mathcal{E}$.}\label{spincomb}
\end{center}
\end{figure}

This paper is organized as follows. In section \ref{mirror} mirror
inversion in spin chains is thoroughly described. In section
\ref{graph} the graph state generation scheme exploiting mirror
inversion is briefly reviewed. Section \ref{character} outlines
the methods we apply to characterize the performance of the EB as
a single-qubit channel and as a two-qubit gate in the presence of
noise. In section \ref{noise} the class of local noise that is
considered in this work is introduced. The influence of these
local noise models are then systematically analyzed in section
\ref{results} for both the single-qubit channel and two-qubit gate
scenario. For the readers convenience the detailed results of
section \ref{results} are summarized in its first subsection. We
then examine the implications of these results for the generation
of a five qubit linear cluster state and
Greenberger-Horne-Zeilinger (GHZ) state in section \ref{graphgen}
before concluding in section \ref{conclusion}.

\section{Mirror-inverting spin chains}
\label{mirror}

Our starting point is a spin-$\half$ chain composed of $N$ spins
which is governed by an XX Hamiltonian of the form (taking $\hbar =
1$)
\begin{eqnarray}
H_S &=&
-\frac{J}{2}\sum_{j=1}^{N-1}t_j(\sigma^{x}_j\sigma^{x}_{j+1} +
\sigma^{y}_{j+1}\sigma^{y}_j) + \frac{1}{2}\sum_{j=1}^N
h_j(\mathbbm{1}-\sigma^{z}_j), \label{HS}
\end{eqnarray}
with spatially dependent spin couplings $t_j$ and local fields
$h_j$. We denote the $\sigma^z$ basis states of the chain as
$\ket{q_1,\dots,q_N}$ with $q_j \in \{0,1\}$ representing $\uparrow$
and $\downarrow$ respectively. Since $[H_S,\mathbbm{N}]=0$, where
$\mathbbm{N}
=\half\sum_{j=0}^N(\mathbbm{1}-\sigma^z_j)$~\footnote{The operator
$\mathbbm{N}$ counts the number of spins which are $\downarrow$.},
then $H_S$ is block-diagonal with respect to subspaces
$\mathcal{H}_\mathbbm{n}$ spanned by states $\ket{q_1,\dots,q_N}$
with $\sum_j q_j = \mathbbm{n}$. The spin chain Hamiltonian $H_S$
can be mapped to a 1D spinless fermionic lattice model using the
Jordan-Wigner transformation (JWT)~\cite{Sachdev-1999} giving
\begin{eqnarray}
H_F &=& -J\sum_{j=1}^{N-1}t_j(c^{\dagger}_jc_{j+1} +
c^{\dagger}_{j+1}c_j) + \sum_{j=1}^N h_jc^{\dagger}_jc_j,
\nonumber
\end{eqnarray}
and subsequently diagonalized into an explicitly free-fermion
bi-linear form
\begin{eqnarray}
H_D &=& \sum_{k=1}^{N}\epsilon_k\,a^{\dagger}_ka_{k},\nonumber
\end{eqnarray}
with spectrum $\epsilon_k$. Here both $c^{\dagger}_j(c_j)$ and
$a^{\dagger}_k(a_k)$ are fermionic creation (annihilation)
operators, obeying the usual anticommutation relation, associated
to lattice site $j$ and the energy eigenstate $k$ respectively.
Under this mapping $\mathbbm{N}=\sum_j c^{\dagger}_jc_j$ and the
subspaces $\mathcal{H}_{\mathbbm{n}}$ it defines are identified
with the fermion number. The fermion vacuum is then $\vac =
\ket{0,\dots,0}$ with energy $E_{\mathrm{vac}}=0$ and spin states
$\ket{q_1,\dots,q_N}$ become $\mathbbm{n}$ fermion Fock states
$\ket{q_1,\dots,q_N} \mapsto
(c^{\dagger}_1)^{q_1}\dots(c^{\dagger}_N)^{q_N}\vac$ with the
operator ordering following the lattice numbering. We denote the
blocks of $H_F$ acting on subspaces $\mathcal{H}_{\mathbbm{n}}$ as
$H_F^{(\mathbbm{n})}$ and since $H_F$ is a non-interacting
Hamiltonian its properties are entirely defined by its
single-particle Hamiltonian $H_F^{(1)}$.

To be mirror inverting all localized states $\ket{j} =
c^{\dagger}_j\vac$ in $\mathcal{H}_1$ are required to evolve after
a given fixed time $\tau$ under $H_F^{(1)}$ into the localized
state $\ket{\bar{j}}$ (up to a phase) where $\bar{j} = N-j+1$ is
the mirror location in the lattice. While this places constraints
on the couplings $t_j$ and fields $h_j$ there are still an
infinite number of permissible
choices~\cite{Yung-PRA-2005,Karbach-PRA-2005}. In this work we
exclusively consider the simplest and fastest mirror inverting
couplings~\cite{Yung-PRA-2006} where $t_j = \half\sqrt{j(N-j)}$
and $h_j = h$. With this choice $H_F^{(1)}$ takes the form
$H_F^{(1)} = -JS_x + h$ where $S_x$ is the $x$-axis angular
momentum operator for a spin-$\mathcal{S}$ pseudo-particle where
$\mathcal{S} = \half(N-1)$. Localized $\mathbbm{n}=1$ states are
then identified with $S_z$ eigenstates $\{\ket{\mathcal{S},l}_z\}$
of the pseudo-spin through the ordering $\ket{1} =
\ket{\mathcal{S},-\mathcal{S}}_z$, $\dots$, $\ket{N} =
\ket{\mathcal{S},\mathcal{S}}_z$. If we now consider the time
evolution in $\mathcal{H}_1$ for a time $\tau=\pi/J$ we see that
$U^{(1)} = \exp(-\i H_F^{(1)}\tau) = \exp(-\i h\pi/J)\exp(\i\pi
S_x)$ is a rotation of the pseudo-spin by $\pi$ about its $x$-axis
and is therefore equivalent to the mirror inversion of a single
fermion in the
lattice~\cite{Christandl-PRL-2004,Christandl-PRA-2005}.
Interestingly, we note that similar effects also take place in a
chain of coupled harmonic oscillators as discussed in
reference~\cite{Plenio-NJP-2004}.

Moving our consideration back to the full state space of the lattice
it follows that the localized modes $c^{\dagger}_j$ are related to
the energy eigenmodes $a^{\dagger}_k$ via irreducible
representations $d_{jk}\left(\frac{\pi}{2}\right)$ of a $\pi/2$
rotation about the $y$-axis of the pseudo-spin~\cite{Chaichian-1998}
as $a^{\dagger}_k = \sum_j
d_{jk}\left(\frac{\pi}{2}\right)c^{\dagger}_j$. The angular momentum
couplings also result in the spectrum of $H_D$ being linear as
$\epsilon_k = J(k - \mathcal{S} - 1) + h$ over the range $\epsilon_k
\in [-\mathcal{S}+h, \mathcal{S}+h]$, and so in order to ensure that
the state $\vac$ is the non-degenerate ground state of the system,
for all $J$, we require $h>\mathcal{S}J$. We can define the
many-body gap between the vacuum ground state and the first excited
state as $\Delta$ giving $h = \mathcal{S}J + \Delta$ and for
$\Delta>0$ the first excited state is always in the $\mathcal{H}_1$
subspace. Note also that with this definition in the limit
$J\rightarrow 0$ we have that $\Delta$ is the local gap for each
decoupled spin. Additionally, we can choose $\Delta/J$ as an even
number such that mirror inversion proceeds with no phase modulo
$2\pi$.

The mirror inverting dynamics in $\mathcal{H}_1$ is equivalent to
the transformation $Uc^{\dagger}_jU^{\dagger} =
c^{\dagger}_{\bar{j}}$ on the localized modes with $U = \exp(-\i
H_F\tau)$. Applying this evolution to an arbitrary $\mathbbm{n}$
fermion Fock state, and performing the inverse JWT, mirror
inversion results in
\begin{eqnarray}
e^{-iH_F\tau}\ket{q_1,\dots,q_N}&=&e^{-i\pi\Sigma_{\mathbbm{n}}}\ket{q_N,\dots,q_1},
\label{fock}
\end{eqnarray}
where $\Sigma_{\mathbbm{n}} = \half\mathbbm{n}(\mathbbm{n}-1)$ is
the number of anti-commutations of the operators $c^{\dagger}_j$
required to reestablish the correct ordering. The simplest
utilization of mirror inversion is state transfer where we
restrict our consideration to the subspace
$\mathcal{H}_0\oplus\mathcal{H}_1$ spanned by the spin-polarized
state $\vac$ and the single spin-flip states $\ket{j}$. We then
encode an input qubit as a superposition $\ket{\psi} = \nu_0\vac +
\nu_1\ket{1}$ using the first spin in the chain and under purely
coherent evolution this state is transferred perfectly to the last
spin as $\ket{\psi} = \nu_0\vac +
\nu_1\ket{N}$~\cite{Christandl-PRL-2004,Christandl-PRA-2005}. The
same conclusion follows trivially for a mixed input state.

\begin{figure}
\begin{center}
\includegraphics[width=11cm]{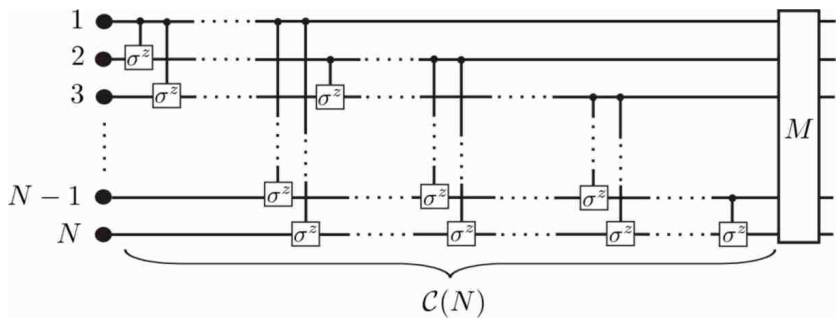}
\caption{The quantum circuit $\mathcal{C}(N)$ composed of
c-$\sigma^z$ gates between all distinct pairs of qubits obtained
by evolving the mirror-inverting spin chain with Hamiltonian $H_S$
for a time $\tau$.}\label{circuitgraph}
\end{center}
\end{figure}

A more general use of mirror inversion follows from noting that
the phase $\pi\Sigma_{\mathbbm{n}}$ in \eqr{fock} is non-linear in
$\mathbbm{n}$ and only appears between subspaces with different
fermion number for $\mathbbm{n} \geq 2$. Thus for input states of
the chain which involve superpositions spanning several
multi-particle subspaces these phases will create entanglement in
the mirror-inverted output
state~\cite{Clark-NJP-2005,Yung-PRA-2005}. More precisely, the
evolution $U$  of the chain for a time $\tau$ is equivalent to a
quantum circuit $\mathcal{C}(N)$ composed of c-$\sigma^z$ gates
between all distinct pairs of $N$ qubits followed by the inversion
operator $M$, as shown in \fir{circuitgraph}. This circuit has the
useful property that if any $N-q$ spins in the chain are in the
state $\ket{0}$, then this circuit reduces to $\mathcal{C}(q)$
between the remaining $q$ qubits, independent of their locations,
followed by the full inversion $M$ of the chain.

\section{Graph state generation with an engineered spin ladder}
\label{graph}

Here we briefly review the scheme given in~\cite{Clark-NJP-2005}
where the general multi-qubit circuit $\mathcal{C}(N)$ implemented
by a mirror inverting chain is exploited to construct graph
states. This is achieved by considering a spin-ladder with a {\em
comb}-like arrangement of couplings as depicted in
\fir{spincomb}(c). One chain of the ladder possess fixed mirror
inverting couplings and forms the EB, while the other chain is
composed of decoupled spins forming the register R. We assume that
spins in the register can be individually manipulated and
measured. Dynamical control of the spin couplings is restricted to
those between adjacent spins in the EB and R where we require the
ability to rapidly implement a swap gate. In this way entanglement
generation is achieved by repeatedly swapping qubits between R and
EB and thereby using the quantum circuit $\mathcal{C}(N)$.

The entire spin ladder is taken to be initialized in a spin
polarized state. The scheme begins by choosing a set of register
spins $G$ that will be the graph qubits, and transforming all of
them to $\ket{+}$. For any subset $Q\subset G$ of graph qubits
which are transferred into the EB and evolved for a time $\tau$
the resulting circuit $\mathcal{C}(|Q|)$ will apply c-$\sigma^z$
gates between all of the corresponding graph vertices. In the case
where two graph qubits in the set $Q$ do not already possess an
edge between them this process will establish one, otherwise it
will remove the edge. By proceeding iteratively we can induce any
pattern of edges between the graph qubits $G$. Starting with
$g=1$, we
\begin{enumerate}
 \item transfer the $g$-th graph qubit from $G$, and all graph qubits
$g_c > g$ which are required to connect to $g$, as specified by
the graphs adjacency matrix $\Gamma$, into the EB;

\item allow the EB to evolve for a time $\tau$ and create a
complete set of connections between all these previously
unconnected vertices;

\item then transfer qubit $g$ back to the register while leaving
the qubits $g_c$ to evolve for one cycle longer in the EB,
subsequently removing all the connections between them;

\item  finally the qubits $g_c$ are transferred back to the
register and step (i) is repeated with $g\mapsto g+1$.
\end{enumerate}
Thus, any graph with $n$ vertices can be generated in at most {\em
O}$(2n)$ uses of the EB in contrast to {\em O}$(n^2)$ steps if the
EB was used to implement single c-$\sigma^z$ gates only. Although
the EB has a linear topology, by using this method any two qubits
in the register can be entangled thereby allowing for arbitrary
topologies of the graph state. To avoid overlap between EB and
register graph qubits after inversion one may choose $|G| \leq
\lceil{N/2}\rceil$ with locations in the first half of the
register.

\section{Characterizing a noisy spin-chain}
\label{character}

The main aim of this work is to characterize the effect of noise
on the performance of the EB and determine its implications for
using the EB within the graph state generation scheme. For
simplicity we determine the performance of the EB at implementing
its two most basic operations, namely acting as a single-qubit
quantum channel and as a mediator of a two-qubit c-$\sigma^z$
gate. These represent the minimal operations required for the EB
to be used for graph state generation. For this reason we consider
the effect of noise only on the EB spin chain and not the register
R. Additionally we focus on the scenario in which the input and
output qubits are the end spins of the EB. Before describing any
specifics about the noise we first outline some general
theoretical tools which provide insightful measures of
performance.

\subsection{Average fidelity}
Suppose we have a system which, when no noise is present, performs
a particular unitary operation $U$. With the inclusion of noise
the action of the system is instead described by a superoperator
$\Lambda$. How close the noisy operation remains to $U$ for a
particular initial pure state
$\ket{\psi}\in\mathbbm{C}^\mathbbm{d}$ can be quantified by the
fidelity~\cite{Nielsen-2000}
\begin{eqnarray}
F(\psi) &=&
\bracket{\psi}{U^{\dagger}\,\Lambda\{\outprod{\psi}{\psi}\}\,U}{\psi}\,.
\label{specificfid}
\end{eqnarray}
The overall performance of the noisy system at implementing $U$ can
then be measured by the average of this fidelity over all possible
initial pure states
\begin{eqnarray}
\av{F} &=& \int_{S^{2\mathbbm{d}-1}} F(\psi)\,\mathrm{d}\psi,
\nonumber
\end{eqnarray}
where integration is over the unit sphere $S^{2\mathbbm{d}-1}$ in
$\mathbbm{C}^\mathbbm{d}$ and $\mathrm{d}\psi$ is the normalized
measure on the sphere, also known as a Haar measure. For the case
of a single qubit this is equivalent to integration over the Bloch
sphere as $\int_{S^3}\,{\rm d}\psi =
\frac{1}{4\pi}\int_{-\pi}^{\pi} {\rm d}\phi\int_0^{\pi}{\rm
d}\theta\,\sin(\theta)$. Now given a Kraus decomposition of the
superoperator $\Lambda$ as
\begin{eqnarray}
\Lambda\{\rho\} &=& \sum_{m=1}^{\mathbbm{d}^2} A_m \rho
A_m^{\dagger}, \nonumber
\end{eqnarray}
where $A_m$ are Kraus operators there is a compact formula for
$\av{F}$ in any dimension $\mathbbm{d}$. Firstly, we form a new
superoperator $\mathcal{E}$ with Kraus operators $E_m = A_m
U^{\dagger}$, such that $\mathcal{E}\{U\rho
U^{\dagger}\}=\Lambda\{\rho\}$, which describes exclusively the
effect of noise. It can then be shown
\cite{Dankert-2005,Pedersen-2007} that
\begin{eqnarray}
\av{F} &=&
\frac{1}{\mathbbm{d}(\mathbbm{d}+1)}\left(\sum_{m=1}^{\mathbbm{d}^2}
|\tr(E_m)|^2 + \mathbbm{d}\right).\label{fidelity}
\end{eqnarray}
We exploit this formula to determine the single qubit channel (or
$\mathbbm{1}$ operation) fidelity, and the gate fidelity for the
effective c-$\sigma^z$ operation between two qubits achieved with
a noisy EB.

\subsection{Entanglement breaking and generation}
While the average fidelity provides a quantitative measure of a
noisy operation, a more qualitative way of characterising the
severity of the noise is to determine whether the corresponding
superoperator $\Lambda^{[1]}$, which acts on one subsystem,
preserves any entanglement that the subsystem has with other
external systems. Quite generally if $\Lambda^{[1]}$ acts on the
subsystem $b$, with Hilbert space $\mathcal{H}_b =
\mathbbm{C}^{\mathbbm{d}_b}$, it is described as {\em entanglement
breaking}~\cite{Hein-PRA-2005} if the final state
$\rho^{\mathrm{out}}_{ab} =
\mathbbm{1}_a\otimes\Lambda^{[1]}_b\{\rho^{\mathrm{in}}_{ab}\}$ is
separable for every (possibly entangled) initial state
$\rho^{\mathrm{in}}_{ab}$ of the composite system of $b$ and
another subsystem $a$ with Hilbert space $\mathcal{H}_a =
\mathbbm{C}^{\mathbbm{d}_a}$. Becoming entanglement breaking
therefore signifies that the channel can no longer be used to
distribute entanglement.

Remarkably, for a single-qubit ($\mathbbm{d}_b=2$) the PPT
criterion~\cite{Peres-PRL-1996,Horodecki-PLA-1996} (see \ref{ppt})
in combination with the Jamiolkowski
isomorphism~\cite{Jamiolkowski-RMP-1972} (see \ref{Jamiolkowski}
and \fir{isomorphism}) give a straightforward condition for
$\Lambda^{[1]}$ to be entanglement breaking. Firstly, it is
sufficient to compute the state $\rho^{\Lambda}$ from the
Jamiolkowski isomorphism (see \fir{isomorphism}(a)), where
$\mathbbm{d}_a=\mathbbm{d}_b=2$, since this contains all the
properties of $\Lambda^{[1]}$. It then follows that
$\Lambda^{[1]}$ is entanglement breaking (for any $\mathbbm{d}_a$)
if and only if the state $\rho^{\Lambda}$ is separable since this
implies that $\Lambda^{[1]}$ has a Kraus representation composed
entirely of projectors. Finally, since $\rho^{\Lambda}$ is a two
qubit state its separability follows directly from the PPT
criterion. The entanglement breaking characteristics of the EB
when acting as a single-qubit channel are of importance since the
graph state generation scheme involves its successive use. We
therefore have a minimum requirement that for the EB to be useful
it must, at the very least, preserve any entanglement that an
input qubit may have with other external qubits, such as those in
the register, when acting purely as a quantum channel. This then
provides an essential, albeit optimistic, bound to its tolerance
for noise.

\begin{figure}
\begin{center}
\includegraphics[width=9cm]{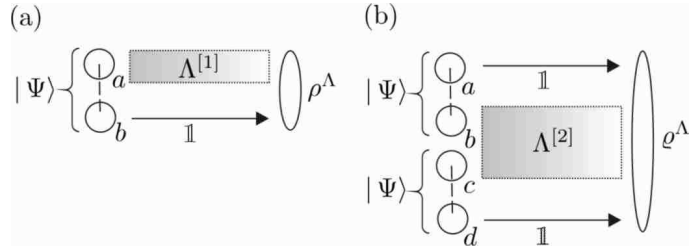}
\caption{A schematic representation of the Jamiolkowski
isomorphism used to characterize a superoperator (a)
$\Lambda^{[1]}$ acting on a single subsystem and (b)
$\Lambda^{[2]}$ acting on a pair of subsystems in the maximally
entangled state $\ket{\Psi}$ (see \ref{Jamiolkowski} for details),
as quantum states $\rho^{\Lambda}$ and $\varrho^{\Lambda}$
respectively.}\label{isomorphism}
\end{center}
\end{figure}

When the evolution of two subsystems is described by a superoperator
$\Lambda^{[2]}$ it is clearly of interest to determine when this
evolution is capable of generating entanglement between these
subsystems~\footnote{When this results in an entangled mixed state
it can then, in principle, be distilled.}. Specifically, for
$\rho^{\mathrm{out}}_{bc} =
\Lambda^{[2]}_{bc}\{\rho^{\mathrm{in}}_{bc}\}$ we may ask when is
$\rho^{\mathrm{out}}_{bc}$ always separable for all possible
separable initial state $\rho^{\mathrm{in}}_{bc}$? This implies that
the superoperator $\Lambda^{[2]}$ never generates entanglement. The
question can be answered by again appealing to the Jamiolkowski
isomorphism via the state $\varrho^{\Lambda}$ associated to
$\Lambda^{[2]}$. It follows that $\Lambda^{[2]}$ is of product form
$\Lambda^{[2]}_{bc} = \Lambda^{[1]}_b\otimes\Lambda^{[1]}_c$ and
incapable of generating entanglement if its corresponding state
$\varrho^{\Lambda}$ is separable with respect to the bipartition of
the system as $(ab)(cd)$ as in \fir{isomorphism}(b). Thus the
property of {\em entanglement generation} can also be phrased as a
state separability problem. For two qubits ($\mathbbm{d}_a =
\mathbbm{d}_b = 2$) the mixed state $\varrho^{\Lambda}$ describes
four qubits. In this case the PPT criterion only provides a
necessary condition for the $(ab)(cd)$ separability of this state.
Thus the PPT criterion can only determine a point at which we can no
longer be certain whether $\Lambda^{[2]}$ can generate entanglement.
Nonetheless this point provides a quantitative cut-off which should
be avoided if the noisy entangling operation is to be of practical
use.

\section{Noise models}
\label{noise} We consider noise which is described by a quantum
master equation of Lindblad form
\begin{eqnarray}
\frac{\partial}{\partial t}\rho(t) &=& -\i[H_S,\rho(t)] +
\mathcal{L}\{\rho(t)\}, \label{fullmaster}
\end{eqnarray}
where $\mathcal{L}\{\cdot\}$ is the Lindbladian describing the
incoherent contribution to the evolution of the density matrix
$\rho(t)$. The microscopic derivation of such a master equation
relies on the Born-Markov approximation and is typically found to
be accurate for systems with a weak coupling to a much larger
environment~\cite{Breuer-2002}.

We consider a subclass of this noise model where each spin
experiences an independent local environment so the Lindbladian
decomposes as a sum
$\mathcal{L}\{\cdot\}=\sum_j\mathcal{L}_j\{\cdot\}$. We make one
further restriction and consider the local Lindbladian
$\mathcal{L}_j\{\cdot\}$ to be of a physically well motivated form
commonly encountered in quantum optical problems after applying the
rotating wave approximation~\cite{Briegel-PRA-1993,Hein-PRA-2005}.
Specifically,
\begin{eqnarray}
\label{master} \mathcal{L}_j\{\rho(t)\} &=&
\frac{\alpha}{2}\,[2\,\sigma_j^-\rho(t)\sigma_j^+
-\sigma_j^+\sigma_j^-\rho(t) -
\rho(t)\sigma_j^+\sigma_j^-]\nonumber \\
&& + \frac{\beta}{2}\,[ 2\,\sigma_j^+\rho(t)\sigma_j^-
-\sigma_j^-\sigma_j^+\rho(t) -
\rho(t)\sigma_j^-\sigma_j^+]\nonumber \\
&& +\frac{\gamma}{2}\,[\,\sigma_j^z\rho(t)\sigma_j^z - \,\rho(t)],
\end{eqnarray}
where $\alpha$, $\beta$ and $\gamma$ are the rates for jumps
$\ket{\uparrow} \rightarrow \ket{\downarrow}$, $\ket{\downarrow}
\rightarrow \ket{\uparrow}$, and pure dephasing, respectively. To
give an overview of the physics contained in this model let us
consider the situation where $J=0$ in $H_S$, defined in \eqr{HS},
so each spin decouples with a local Hamiltonian of the form
$H_j=\frac{\Delta}{2}(\mathbbm{1}-\sigma_j^z)$. If we move to the
interaction picture of $H_j$ for each spin~\footnote{We shall
denote the interaction picture of a Hamiltonian $H$ by a tilde as
$\tilde{O} = e^{\i Ht}\,O\,e^{-\i Ht}$.} it is readily found that
the evolution of the $j$-th spin is described by $\tilde{\rho}(t)
= e^{\mathcal{L}_jt}\{\tilde{\rho}\}$ with $\mathcal{L}_j$
remaining in terms of the untransformed operators in \eqr{master}
due to phase cancellation. We now write the initial state
$\tilde{\rho}$ as
\begin{eqnarray}
\tilde{\rho} &=&
\half\left[\mathbbm{1}+\av{\tilde{\sigma}_x}\tilde{\sigma}_x+\av{\tilde{\sigma}_y}\tilde{\sigma}_y+\av{\sigma_z}\sigma_z\right],\nonumber
\end{eqnarray}
from which the general solution is found to
be~\cite{Briegel-PRA-1993}
\begin{eqnarray}
e^{\mathcal{L}t}\{\rho\} &=& \half\left[\mathbbm{1} +
\av{\sigma_z}_s\,\sigma_z +e^{-\half(\alpha + \beta +
2\gamma)t}(\av{\tilde{\sigma}_x}\tilde{\sigma}_x+\av{\tilde{\sigma}_y}\tilde{\sigma}_y)
\right.
\nonumber \\
&& \left. +~e^{-(\alpha +
\beta)t}(\av{\sigma_z}-\av{\sigma_z}_s)\sigma_z\right].\nonumber
\end{eqnarray}
As is well known this solution shows exponential convergence with
rate $\alpha + \beta$ of $\av{\sigma_z(t)}$ to its stationary ($t
\rightarrow \infty$) value of $\av{\sigma_z}_s =\frac{\beta -
\alpha}{\alpha +\beta}$ and the exponential decay, with rate
$\half(\alpha + \beta + 2\gamma)$, of the coherences
$\av{\tilde{\sigma}_{\pm}(t)}$ to their stationary value
$\av{\sigma_{\pm}}_s = 0$. The general solution to this noise
model can be expressed in a Kraus form~\cite{Nielsen-2000} with
Kraus operators
\begin{eqnarray}
\label{kraus} E_{1} &=& \left(\begin{array}{cc}
\Upsilon_1 & 0  \\
0 & \Upsilon_2
\end{array}\right),\quad E_{2} = \sqrt{P_{\uparrow}}\left(\begin{array}{cc}
0 & \sqrt{1-e^{-(\alpha + \beta)t}}  \\
0 & 0
\end{array}\right), \nonumber \\
E_{3} &=& \left(\begin{array}{cc}
\Upsilon_3 & 0 \\
0 & \Upsilon_4 \end{array}\right), \quad E_{4} =
\sqrt{P_{\downarrow}}\left(\begin{array}{cc}
0 & 0  \\
\sqrt{1-e^{-(\alpha + \beta)t}} & 0 \end{array}\right),
\end{eqnarray}
where $P_{\uparrow} = \frac{\beta}{\alpha+\beta}$, $P_{\downarrow}
= \frac{\alpha}{\alpha+\beta}$ are the stationary spin populations
and $\Upsilon_i$ are functions of $\alpha,\beta$ and $\gamma$
which we give explicitly in \ref{limits}. We also show in
\ref{limits} that this noise model reduces to a number of well
known and simpler models in specific limits. In particular, if we
parameterize the rates as (taking Boltzmann's constant $k_B=1$)
\begin{eqnarray}
\alpha(T) &=& \kappa\,\frac{e^{-\Delta/T}}{(1+e^{-\Delta/T})}
\quad \mathrm{and} \quad \beta(T) =
\kappa\frac{1}{(1+e^{-\Delta/T})} \label{thermalrates}
\end{eqnarray}
for an arbitrary $\gamma$, we obtain a total decay rate $\alpha(T)
+ \beta(T) = \kappa$ that is independent of $T$. In this case the
stationary density matrix $\rho_s = \lim_{t\rightarrow\infty}
\rho(t)$ for the spin is equivalent to a thermal state of
temperature $T$. Consequently in this regime the master equation
describes {\em finite-temperature} noise caused by the coupling to
a generic thermal reservoir at temperature $T$ local to each spin.
At $T=0$ and $\gamma=0$ we have $\alpha = 0$ and $\beta=\kappa$
which describes {\em decay} noise. For $T\rightarrow\infty$ we
have $\alpha = \beta = \half \kappa$ and after setting
$\gamma=\half\kappa$, so the populations and coherences decay at
the same rate, this results in {\em depolarizing} noise. Taking
$\alpha=\beta=0$ and $\gamma=\kappa$ we obtain pure {\em
dephasing} noise. We shall consider each of these limiting cases
as local noise in the EB.

If we now consider this class of noise in the context of a
single-qubit channel we can determine the properties which were
outlined in section \ref{character}. Indeed using the Kraus
operators in \eqr{kraus} the average fidelity of the channel can be
computed via \eqr{fidelity} and is given in full in \ref{limits}.
The channel can be shown to become entanglement breaking if and only
if the following condition is satisfied~\cite{Hein-PRA-2005}
\begin{eqnarray}
2P_{\uparrow}P_{\downarrow}\,e^{2\gamma
t}\{\cosh([\alpha+\beta]t)-1\} &\geq& 1. \label{critical}
\end{eqnarray}
In particular this result immediately indicates that, regardless
of $\gamma$, whenever $\alpha=0$ or $\beta=0$ the channel is never
entanglement breaking for finite coupling $\kappa$ and times $t$
since either $P_{\downarrow}=0$ or $P_{\uparrow}=0$, respectively.
This result similarly holds when both $\alpha=0$ and $\beta=0$,
giving a pure dephasing channel~\footnote{Note in the case of pure
dephasing the definitions of $P_{\uparrow}$ and $P_{\downarrow}$
in terms of $\alpha$ and $\beta$ are meaningless and the
stationary populations follow from the arbitrary initial state.},
since we have $\cosh([\alpha+\beta]t)-1=0$. In contrast a finite
temperature channel (for any $\gamma$) can always become
entanglement breaking for a finite $\kappa$ and $t$. For
$\gamma=0$ this entanglement breaking occurs for a coupling
\begin{eqnarray}
\kappa_c &\geq&
\frac{J}{\pi}\cosh^{-1}\left[\frac{(1+e^{-\Delta/T})^2}{2\,e^{-\Delta/T}}
+ 1\right], \label{thermalcrit}
\end{eqnarray}
taking $t = \tau$, and this saturates at $\kappa_c \approx
0.56\,J$ for $T \rightarrow \infty$. The presence of dephasing
reduces this threshold. An important special case is where
$\gamma=\half\kappa$ which in the $T \rightarrow \infty$ limit
gives a depolarizing channel with a threshold $\kappa_c \geq
\frac{J}{\pi}\log(3)\approx 0.35\,J$.

\section{Numerical method}
The numerical calculations we perform in this work is restricted
to the class of 1D quantum lattice systems described by a master
equation which include a Hamiltonian and a Lindbladian that are
both composed of terms involving at most nearest-neighboring
sites. It can be seen that both $H_S$ and $\mathcal{L}$ introduced
in section \ref{noise} satisfy this constraint. The real time
evolution for this class of master equation can be computed
efficiently and to near-exact precision for systems composed of
many sites using the mixed-state version of the Time Evolving
Block Decimation (TEBD)
algorithm~\cite{Zwolak-PRL-2004,Vidal-PRL-2003,Vidal-PRL-2004}. We
refer the reader to the literature for a detailed description of
this method and note here only that for the calculations presented
we found that a truncation parameter~\cite{Vidal-PRL-2004} up to
$\chi=20$ was sufficient. In \fir{calcsetup} we depict the type of
numerical calculation we have performed with this algorithm. These
are based on the Jamiolkowski isomorphism which for two qubits
requires the spin chain to be initialized in a pure state
$\ket{\Psi}\otimes\ket{0\cdots0}\otimes\ket{\Psi}$ with the two
end spins being in a maximally entangled state
$\ket{\Psi}=(\ket{00} + \ket{11})/\sqrt{2}$ with corresponding
ancillary spins shown in \fir{calcsetup}(a). We then use TEBD to
time evolve the spins $1,\cdots,N$ in the chain in the presence of
noise and finally compute the reduced density matrix
$\varrho^{\Lambda}$ of the two ancillary spins and the two end
spins of the chain as depicted in \fir{calcsetup}(b). The state
$\varrho^{\Lambda}$ then completely characterizes the accumulative
noisy operation of the chain $\Lambda^{[2]}$ for two qubits.

\begin{figure}
\begin{center}
\includegraphics[width=6cm]{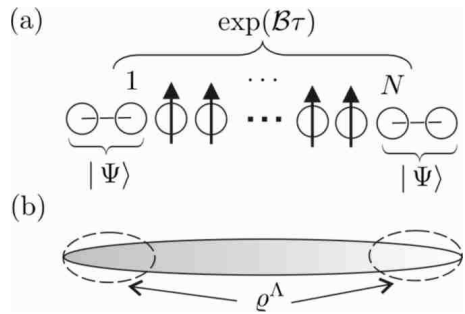}
\caption{The setup used in numerical calculations to determine the
effective two qubit superoperator $\Lambda^{[2]}$ of a noisy
mirror-inverting chain. (a) Following the Jamiolkowski isomorphism
the initial state $\rho$ is a spin-polarized chain
$\ket{0\cdots0}$ aside from the end spins which are in maximally
entangled states $\ket{\Psi}$ with corresponding ancillary spins.
The spins $1,\cdots,N$ in the chain are then evolved for a time
$\tau$ while being exposed to noise. The total dynamical evolution
of the chain is then described by the superoperator
$\exp(\mathcal{B}\tau)$ which is the formal solution to
\eqr{fullmaster}. (b) The state $\varrho^{\Lambda}$ corresponding
to $\Lambda^{[2]}$ is then extracted from the overall final state
$\exp(\mathcal{B}\tau)\{\rho\}$ by tracing out all but the end
spin pairs.}\label{calcsetup}
\end{center}
\end{figure}

\section{Results}
\label{results}

\subsection{Summary of results}
Having introduced all the necessary concepts we now investigate
the influence of local decay, dephasing, thermal and depolarizing
noise on the performance of the EB spin chain. We distinguish
between two scenarios, namely where only one qubit is transferred
into the spin chain so it acts as a quantum channel, and where two
qubits are swapped into the chain such that the mirror inversion
performs a c-$\sigma^z$ gate. We summarise the main results here
and refer the reader to the proceeding subsections for more
details.

When used as a single-qubit channel we find that neither local
decay or dephasing noise become entanglement breaking. For local
decay noise we find that the accumulative noise $\Lambda^{[1]}$ of
the chain is identical to the local decay noise on any single spin
and is therefore entirely independent on the chain length $N$. We
show this useful property is a consequence of commuting coherent
and dissipative contributions to the dynamics which is unique to
local decay noise. The EB chain is found to be the most robust to
local decay noise and is able to maintain $\av{F}\geq 0.99$ for
state transfer when $\kappa/J \leq 9.7 \times 10^{-3}$. The case
of local dephasing noise is shown to be well modelled by a quantum
channel subject to the same local dephasing noise along with a
length dependent decay noise. For spin chains of lengths up to
$N=50$ the average fidelity remains above 99\% as long as the
dephasing noise fulfils $\kappa/J \leq 5 \times 10^{-3}$.

In contrast to these types of noise we find that local
depolarizing and thermal noise become entanglement breaking for
certain parameter regimes, for which analytical estimates are
given. We further find that the length independence observed for
the $T=0$ local decay noise persists as a very weak length
dependence for significant non-zero temperatures $T \leq 0.2
\Delta$. For the local depolarizing noise we find that the
critical coupling at which entanglement breaking occurs for a
given chain length $N$ is described by a power law $\kappa_c/J
\approx N^{-x}$ with $x=0.68$. This behaviour appears to be a
consequence of the competition between the speed and spreading of
a spin-packet in the chain. As expected the chain is most severely
affected by local depolarizing noise with $\av{F}\geq 0.99$ only
for $\kappa/J < 3 \times 10^{-4}$ and lengths up to $N=50$, which
is more than an order of magnitude smaller than for local decay or
dephasing noise.

For the two-qubit case we find that the average gate fidelity with
local decay noise does not depend on the length $N$ and remains
above 99\% for couplings below $\kappa/J \leq 4 \times 10^{-3}$.
For thermal noise we find that $\av{F}$ only depends very weakly
on $N$ as long as $T<0.2 \Delta$, as was the case a single-qubit.
However, in contrast to the single-qubit case, this length
independence in $\av{F}$ does not extend to the accumulative noise
$\Lambda^{[2]}$ superoperator itself. Both local dephasing and
depolarizing noise have average gate fidelities which depend on
$N$ and for up to $N=12$ spins the coupling is restricted to
$\kappa/J \leq 2.5 \times 10^{-3}$ and $\kappa/J \leq 4 \times
10^{-4}$ in order for $\av{F}>0.99$. This again indicates that
local depolarizing noise has the most severe influence and
explains why its accumulative noise is well approximated by
product noise.

\subsection{Single-qubit channel}
In this section we consider the EB as a single-qubit channel and
systematically compute the average channel fidelity $\av{F}$ and
the minimum eigenvalue $\epsilon_{\mathrm{min}}$ of the partial
transposition of the mixed state $\rho^{\Lambda}$ that is
isomorphic to accumulative noise of the chain $\Lambda^{[1]}$.
From the PPT criterion this noise is entanglement breaking
whenever $\epsilon_{\mathrm{min}}>0$. We also use the behaviour of
$\av{F}$ over a wide parameter range to fit the noise
$\Lambda^{[1]}$ of the chain to the specific class of single-spin
noise introduced in section \ref{noise} and find that such fits
are possible to very good accuracy.

\subsubsection{Decay noise (low-$T$ limit) -}
\label{decaynoise} As was noted in reference \cite{Cai-PRA-2006},
we find from our numerics that $\av{F}$ displays no dependence on
the length of the chain. Here we show that this unexpected feature
is in fact a consequence of a much stronger result; specifically,
the superoperator $\Lambda^{[1]}$ itself, which characterises the
accumulative noise of the chain, is independent of $N$. This
result implies that $\Lambda^{[1]}$ for any $N$ is equivalent to
$\Lambda^{[1]}$ for a chain with $N=1$. Since a $N=1$ chain is a
single-qubit decay channel, this allows us to conclude that
state-transfer in a mirror-inverting chain with local decay noise
is never entanglement breaking. Additionally, the coupling at
which the fidelity drops to $\av{F}<0.99$ is found to be
$\kappa_f/J=9.7\times 10^{-3}$, independent of $N$.

We now explain the independence of chain length observed. To begin
we take a spin chain composed of $N$ spins and a general
Hamiltonian $H_s$ satisfying $[H_s,\mathbbm{N}]=0$ so $H_s$ is
block diagonal with blocks $H^{(\mathbbm{n})}_s$ in each subspace
$\mathcal{H}_{\mathbbm{n}}$~\footnote{The Hamiltonian $H_S$
defined in section \ref{mirror} is one such example.}. Then we
restrict our considerations to initial states of the chain
$\rho(0)$ whose support is entirely contained in the subspace
$\mathcal{H}_{0}\oplus \mathcal{H}_1$. Evolution due to $H_s$ and
local decay noise has the convenient feature that the support of
the state $\rho(t)$ at any time will also remain entirely within
$\mathcal{H}_{0}\oplus \mathcal{H}_1$. Consequently, we may
project the full master equation of the chain into this subspace
yielding
\begin{eqnarray}
\frac{\partial}{\partial t}\rho(t) &=& -\i[H^{(0)}_s \oplus
H^{(1)}_s,\rho(t)] +
\frac{\kappa}{2}\,\left(2\,P_1(t)\mathbbm{P}_0 -
\mathbbm{P}_1\rho(t) - \rho(t)\mathbbm{P}_1\right).\nonumber
\end{eqnarray}
Here $P_1(t)=\tr(\mathbbm{P}_1\rho(t)\mathbbm{P}_1)$ is the
probability of being in the $\mathcal{H}_1$ subspace, with
$\mathbbm{P}_0$ and $\mathbbm{P}_1$ being the projectors onto the
subspaces $\mathcal{H}_0$ and $\mathcal{H}_1$, respectively. If
this projected master equation is expressed as $\dot{\rho}(t) =
{\tt H}\{\rho(t)\} + {\tt L}\{\rho(t)\}$, where $\tt H$ and $\tt
L$ are the coherent and dissipative superoperators, it follows
that $[{\tt H},{\tt L}]=0$ since $\tt L$ is composed entirely of
projectors onto the same subspaces over which $H_s$ is
block-diagonal. The crucial effect of this commutivity is that
\begin{eqnarray}
\rho(t) = e^{{\tt H}t + {\tt L}t}\{\rho(0)\} &=& e^{{\tt
H}t}e^{{\tt L}t}\{\rho(0)\} = e^{{\tt L}t}e^{{\tt
H}t}\{\rho(0)\}\nonumber.
\end{eqnarray}
Hence the coherent and dissipative contributions to the evolution
are independent and can be applied separately.

In the special case where $H_s=H_S$ is a mirror-inverting
Hamiltonian this property manifests itself directly in the
accumulative noise $\Lambda^{[1]}$. Using the chain as a channel
involves initializing it in a spin-polarized state
\begin{eqnarray}
\rho(0) &=&
\outprod{\uparrow}{\uparrow}_1\otimes\cdots\otimes\varsigma_j\otimes\cdots\otimes\outprod{\uparrow}{\uparrow}_{\bar{j}}\otimes\cdots\otimes\outprod{\uparrow}{\uparrow}_N\nonumber,
\end{eqnarray}
aside from the spin $j$ which is in input state $\varsigma$. If we
first apply the coherent evolution for a time $\tau$ then, as
outlined in section \ref{mirror}, the state becomes
\begin{eqnarray}
\rho_1 = e^{{\tt H}\tau}\{\rho(0)\} &=&
\outprod{\uparrow}{\uparrow}_1\otimes\cdots\otimes\outprod{\uparrow}{\uparrow}_{j}\otimes\cdots\otimes\varsigma_{\bar{j}}\otimes\cdots\otimes\outprod{\uparrow}{\uparrow}_N\nonumber,
\end{eqnarray}
where the state $\varsigma$ is transferred to the mirror spin
$\bar{j}$. Since the whole chain is spin-polarized, aside from at
spin $\bar{j}$, the action of the superoperator $\tt L$ on such a
state is completely equivalent to single-spin decay noise at that
spin alone. Thus, the final output state $\varphi$ of spin
$\bar{j}$ after tracing out all other spins (denoted as $c$)
\begin{eqnarray}
\varphi_{\bar{j}} &=& \tr_{c}(e^{{\tt L}\tau}\{\rho_1\})
\nonumber,
\end{eqnarray}
is identical, irrespective of $N$~\footnote{On the proviso that
the inversion time $\tau$ is kept constant with $N$.}, to the
output state for a chain with $N=1$ where the input state
$\varsigma$ is simply exposed to single spin decay noise for a
time $\tau$.

\subsubsection{Dephasing noise -}
\label{dephaseone} We find that the behavior of
$\epsilon_{\mathrm{min}}$ and $\av{F}$, displayed in
\fir{fig:dephase}(a)-(b), for the accumulative noise of the spin
chain with local dephasing rapidly converges with the length of
the chain. In particular \fir{fig:dephase}(a) shows that the chain
does not become entanglement breaking for the wide range of chain
lengths $N$ and couplings $\kappa/J$ investigated. In
\fir{fig:dephase}(c) we plot the coupling $\kappa_f/J$ at which
the fidelity drops below $\av{F}<0.99$. This plot indicates that
the coupling must not exceed $\kappa/J=5\times 10^{-3}$ for chain
lengths of order $N=50$ for useful fidelities to be achieved.

\begin{figure}
\begin{center}
\includegraphics[width=11cm]{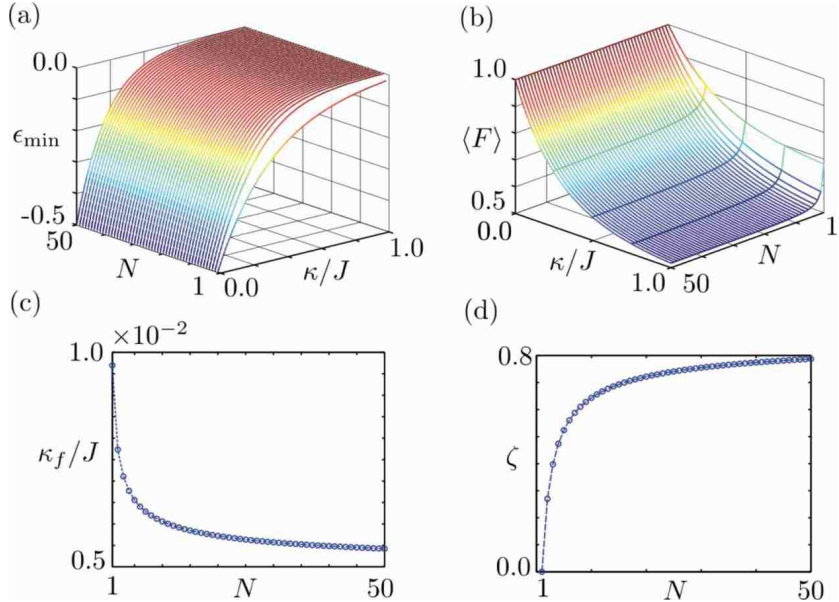}
\caption{For local dephasing noise - (a) The minimum eigenvalue
$\epsilon_{\mathrm{min}}$ of the partial transposition of
$\rho^{\Lambda}$ and (b) the average fidelity $\av{F}$, both as a
function of the chain length $N$ and coupling strength $\kappa/J$.
(c) The coupling $\kappa_f/J$ at which the fidelity shown in (b)
drops below $\av{F}<0.99$ against the chain length $N$. (d) The
fit parameter $\zeta$ as a function of $N$.}\label{fig:dephase}
\end{center}
\end{figure}

One might expect that the average fidelity $\av{F}$ can be
reproduced by assuming that a single qubit is sent through a
purely dephasing channel with a coupling $\kappa$ dependent on
$N$. However, our numerical calculations show this not to be the
case. Instead, we find that $\av{F}$ is fitted extremely well by
assuming that the overall noise $\Lambda^{[1]}$ is simultaneously
decay and dephasing. This model noise is also never entanglement
breaking for finite $\kappa/J$. Using the general expression for
the noise model in \ref{limits} we fitted $\av{F}$ for each $N$ to
the model noise fidelity with $\gamma = \kappa$ and $\beta = \zeta
\kappa$, where $\zeta$ is the only fit parameter. The parameter
$\zeta$ obtained as a function of $N$ is plotted in
\fir{fig:dephase}(d). It shows that as the chain length increases
the decay rate $\beta$ increases and becomes of the same order as
the dephasing rate $\gamma$. Intuitively this type of model might
be expected to describe the accumulative noise of the chain. Like
local decay noise in the section \ref{decaynoise} an initial state
with support in $\mathcal{H}_0\oplus\mathcal{H}_1$ will remain so
at all times. However, unlike local decay noise the coherent and
dissipative contributions to the projected master equation do not
commute and so perfect mirror inversion is not obtained for any
$\kappa/J >0$. As a result the input state on the first spin is
never perfectly refocussed to the $N$-th spin causing a `leakage'
of the $\downarrow$ population over other spins in the chain.
Since the $N$-th spin is the output qubit at time $\tau$ this
effect appears as decay noise.

To gauge how accurate the assumed noise model was compared to the
actual noise superoperator $\Lambda^{[1]}$ we computed the
fidelity~\cite{Nielsen-2000}
\begin{eqnarray}
F_{\Lambda}(\rho^{\Lambda},\rho^{\Lambda}_m) &=&
\tr\left(\sqrt{\sqrt{\rho^{\Lambda}}\rho^{\Lambda}_m\sqrt{\rho^{\Lambda}}}\right)
\label{isofid}
\end{eqnarray}
between the states $\rho^{\Lambda}$ and $\rho^{\Lambda}_m$
isomorphic to $\Lambda^{[1]}$ and the model noise, respectively
(see \ref{Jamiolkowski} for details). We find that over all
parameters considered the infidelity $1-F_{\Lambda} < 3.2\times
10^{-2}$ which indicates that the model is capturing the
accumulative noise of the chain to good approximation.

\subsubsection{Thermal noise (finite-$T$) -}
\label{tempone} For local thermal noise we restrict our
consideration to a suitably weak coupling $\kappa/J = 0.02$ so
that the corresponding average fidelity at $T=0$ is $\av{F} =
0.98$ and still sizable. In \fir{fig:temp}(a)
$\epsilon_{\mathrm{min}}$ is plotted and demonstrates that up to
temperatures $T = \Delta$ and chain lengths $N=50$ the
accumulative noise of the chain is not entanglement breaking. The
behaviour of both $\epsilon_{\mathrm{min}}$ and $\av{F}$ in
\fir{fig:temp}(b) indicates that their insensitivity to the chain
length $N$, seen earlier for the $T=0$ decay noise, persists for
temperatures $T<0.2\Delta$. This is further confirmed by
\fir{fig:temp}(c) where the temperature $T_f/\Delta$ at which the
fidelity drops to 99\% of its value at $T=0$ is above
$T=0.2\Delta$ for chain lengths up to $N=50$.

\begin{figure}
\begin{center}
\includegraphics[width=11cm]{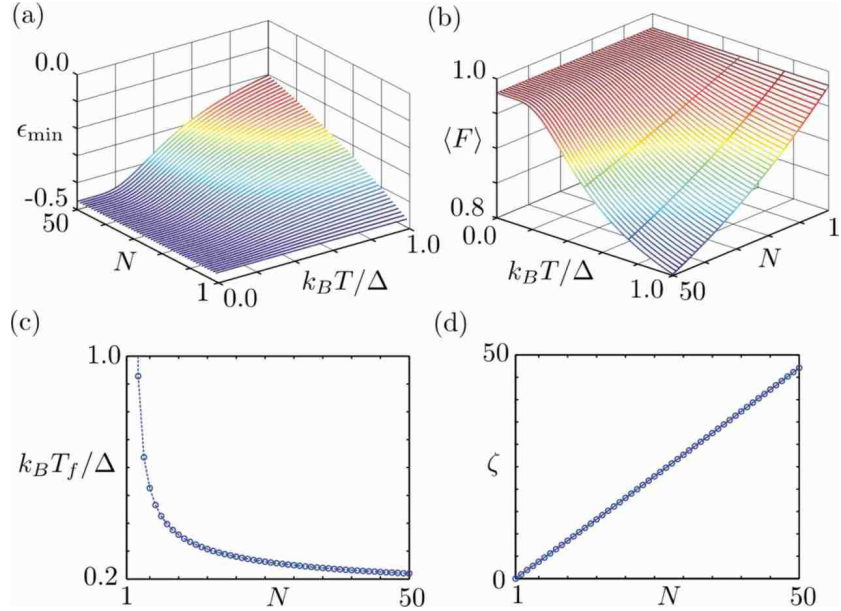}
\caption{For local thermal reservoirs - (a) The minimum eigenvalue
$\epsilon_{\mathrm{min}}$ of the partial transposition of
$\rho^{\Lambda}$ and (b) the average fidelity $\av{F}$, both as a
function of the chain length $N$ and temperature $T$ for noise
caused by a coupling strength $\kappa/J = 0.02$. (c) The
temperature $T_f/\Delta$ at which the fidelity shown in (b) drops
below 99\% of its value at $T=0$ against the chain length $N$. (d)
The fit parameter $\zeta$ as a function of $N$.}\label{fig:temp}
\end{center}
\end{figure}

In order to reproduce the fidelity surface of \fir{fig:temp}(b) we
fitted a noise model in which $\alpha(T)$ and $\beta(T)$ remain
unchanged from those in \eqr{thermalrates} but now include a
non-zero, temperature-dependent dephasing rate $\gamma(T) =
\zeta\alpha(T)$. As a result the effective noise of the chain is
still described by a coupling to thermal reservoir of temperature
$T$. The corresponding fit parameter $\zeta$ for each $N$ is shown
in \fir{fig:temp}(d) and is seen to be linear and very nearly
$\zeta(N) = N$. The accuracy of this model noise compared to the
numerically determined noise was found to be extremely good with
$F_{\Lambda}$ computed via \eqr{isofid} giving $1-F_{\Lambda} <
3.6\times 10^{-5}$.

In section \ref{noise} we found from \eqr{thermalcrit} that for
single-qubit thermal noise, with $\gamma=0$ and acting for a time
$\tau$, a coupling $\kappa/J
> 0.56$ was required for the channel to become entanglement breaking
at $T\rightarrow\infty$. Consequently, a single-qubit channel,
with the weak coupling $\kappa/J = 0.02$ chosen, never becomes
entanglement breaking at any temperature. Using our model for the
accumulative noise of the chain in which a non-zero $\gamma(T)
\approx N\alpha(T)$ emerges we have determined an approximate
analytical expression for the critical length $N_c$ at which
entanglement breaking will occur for a given temperature $T$ and
local coupling $\kappa/J$ as
\begin{eqnarray}
N_c\left(T,\frac{\kappa}{J}\right) \approx
\frac{J(1+e^{-\Delta/T})}{2\kappa\pi
e^{-\Delta/T}}\log\left\{\frac{(1+e^{-\Delta/T})^2}{2\,e^{-\Delta/T}\left[\cosh\left(\frac{\kappa\pi}{J}\right)-1\right]}\right\}
\nonumber.
\end{eqnarray}
For any $\kappa>0$ this function monotonically increases with
decreasing $T$ from a finite asymptotic value at
$T\rightarrow\infty$ and diverges at $T=0$. For the weak coupling
used in this section the critical length at $T\rightarrow\infty$
is $N_c = 111$ spins.

\subsubsection{Depolarizing noise (high-$T$ limit) -}
\begin{figure}
\begin{center}
\includegraphics[width=11cm]{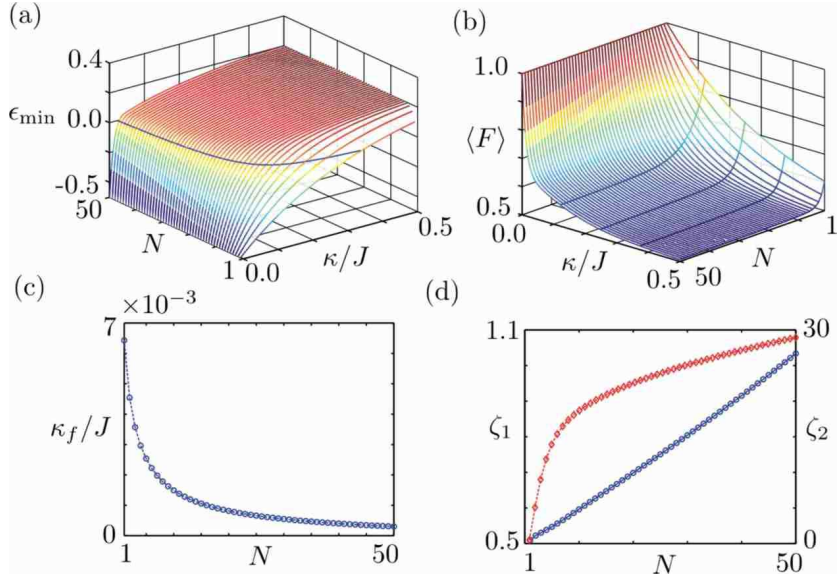}
\caption{For local depolarizing noise - (a) The minimum eigenvalue
$\epsilon_{\mathrm{min}}$ of the partial transposition of
$\rho^{\Lambda}$ and (b) the average fidelity $\av{F}$, both as a
function of the chain length $N$ and coupling strength $\kappa/J$.
The critical coupling $\kappa_c/J$ for each chain length $N$ at
which the accumulative noise is entanglement breaking is given by
the intersection with the $\epsilon_{\mathrm{min}}=0$ plane shown
in (a). (c) The coupling $\kappa_f/J$ at which the fidelity shown
in (b) drops below $\av{F}<0.99$ against the chain length $N$. (d)
The fit parameters $\zeta_1$ (left axis and `$\diamond$') and
$\zeta_2$ (right axis and `$\circ$') as a function of
$N$.}\label{fig:depolar}
\end{center}
\end{figure}

For local depolarizing noise we find that the accumulative noise
of the chain becomes entanglement breaking at a threshold coupling
$\kappa_c/J$ that reduces with the chain length $N$ as shown in
\fir{fig:depolar}(a). The fidelity $\av{F}$ shown in
\fir{fig:depolar}(b) decreases rapidly with $\kappa/J$ for $N>5$.
The coupling $\kappa_f/J$ at which the fidelity drops to
$\av{F}=0.99$ is plotted in \fir{fig:depolar}(c) and indicates
that the condition $\kappa/J<3\times 10^{-4}$ has to be fulfilled
in order to achieve reasonable fidelities for chain lengths up to
$N=50$. Figure \ref{fig:depolar}(b) also shows that the decay of
$\av{F}$ with $\kappa/J$ changes from an exponential behaviour for
small $N$ to a double-exponential behaviour for large $N$. This
indicates that the accumulative noise of the chain does not remain
purely depolarizing, however, we do find that it is still well
approximated by the class of noise introduced in section
\ref{noise}. Using the noise model with
$\alpha=\beta=\zeta_1\kappa$ and $\gamma=\zeta_2\kappa$, and
fitting $\zeta_1$ and $\zeta_2$, the fidelity curves can be
accurately reproduced for all parameters considered. By
restricting $\alpha = \beta$ this model is still thermal noise in
the limit $T \to \infty$, but importantly we allow the total decay
rate to increase from $\kappa$ and also independently allow the
dephasing rate to increase from $\half\kappa$, both as a function
of $N$. The fitting parameters plotted in \fir{fig:depolar}(d)
show that $\zeta_1$ increases from its initial value of $\half$ to
a little over unity, whereas $\zeta_2$ displays a linear increase
with $N$ becoming nearly 30 times larger than $\zeta_1$ for
$N=50$. To establish the validity of this model we compare it to
the actual noise computed numerically by calculating $F_{\Lambda}$
via equation (11). We find that $1-F_{\Lambda} < 1.5\times
10^{-2}$.

The dependence of the critical coupling $\kappa_c/J$ with $N$ is
plotted in \fir{fig:powerlaw} and appears to be described well by
a power-law $\kappa_c/J \approx N^{-x}$ with $x=0.68$. Using the
fitted noise model the critical coupling can be obtained by
solving a special case of \eqr{critical} of the form
   \begin{eqnarray}
\left[\zeta_1(N)\right]^2\,\exp\left[2\zeta_2(N)\pi\
frac{\kappa}{J}\right]\left\{\cosh\left[2\zeta_1(N)\pi
\frac{\kappa}{J}\right]-1\right\}
&\geq& \half, \nonumber
\end{eqnarray}
as a function of $N$ using the functions $\zeta_1(N)$ and
$\zeta_2(N)$ plotted in \fir{fig:depolar}(d). The result of this is
also shown in \fir{fig:powerlaw} and is consistent with a power law
with exponent $x=0.72$.

\begin{figure}
\begin{center}
\includegraphics[width=10cm]{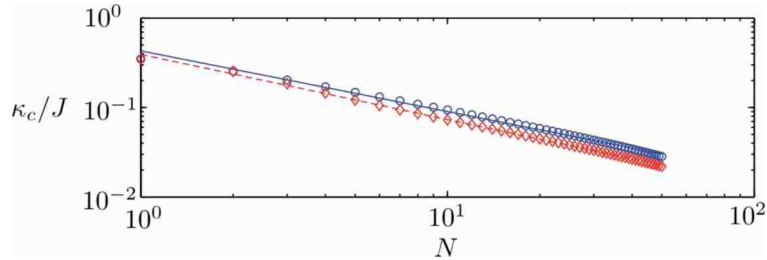}
\caption{The critical coupling $\kappa_c/J$ at which the
accumulative noise of the chain becomes entanglement breaking as a
function of the chain length $N$ on a log-log scale. The numerical
data is plotted with `$\circ$' and the fitted power law
$\kappa_c/J \approx N^{-x}$ is the solid line with an exponent
$x=0.68$. The solution for the critical coupling of the noise
model are plotted with `$\diamond$' and the dotted line is the
power law fit with an exponent $x=0.72$.}\label{fig:powerlaw}
\end{center}
\end{figure}

To gain a qualitative understanding of the origin of this power-law
scaling of $\kappa_c/J$ with $N$ we consider a simple model of this
noise scenario. Specifically we replace each spin of the chain by a
depolarizing channel which preserves the input state with
probability $p_j$ and where state transfer corresponds to the
concatenation of these channels shown in \fir{fig:depolarmodel}(a).
The accumulative noise of this sequence of single qubit channels is
then also a depolarizing channel with probability $p = \prod_j p_j$.
In \fir{fig:depolarmodel}(b) the progression of a spin packet in the
chain resulting from a spin-flip excitation at the first spin is
shown for a sequence of times for $N=50$. With this in mind we take
the probabilities for each channel as $p_j=\exp(-\kappa f_j s_j
\tau)$ where $f_j$ is the fraction of the inversion time $\tau$ the
centre of the spin packet spends at spin $j$, and $s_j$ is the
number of sites the spin packet spreads across when it is in the
region of spin $j$. Both these quantities can be readily derived
from the properties of angular momentum and are plotted in
\fir{fig:depolarmodel}(c). From this we see that the spin packet is
narrow and slow at the edges while being wide and fast at the
centre. The critical coupling for this simple model can be extracted
from \eqr{critical} as
\begin{eqnarray}
\frac{\kappa_c}{J} &=& \frac{\log(3)}{\pi\sum_{j=1}^Nf_js_j}.
\nonumber
\end{eqnarray}
We find that the competition between the spreading and speed of a
spin packet across the chain as a function of its length $N$
naturally gives rise to a power-law dependence for $\kappa_c/J$.

\begin{figure}
\begin{center}
\includegraphics[width=10cm]{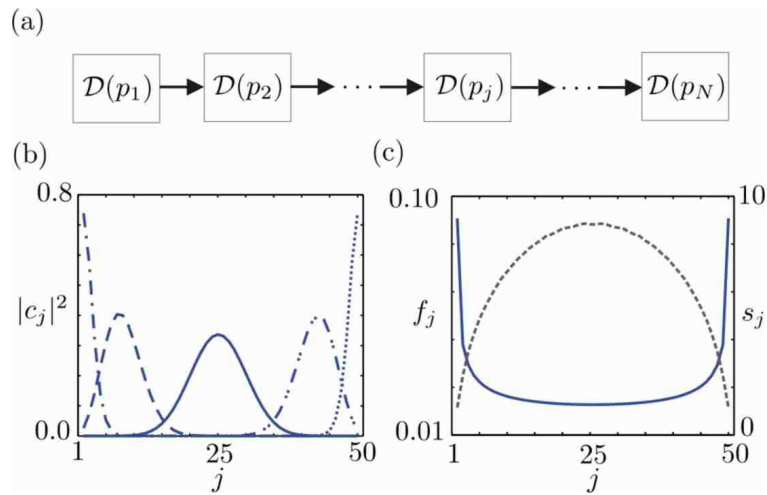}
\caption{(a) A simple model for state transfer in a mirror
inverting chain with local depolarizing noise. Each spin on the
chain is considered to be a depolarizing channel
$\mathcal{D}_j\{\rho\}=p_j\rho + \half(1-p_j)\mathbbm{1}$ with
$p_j=\exp(-\kappa f_j s_j \tau)$. The fraction of the total time
$\tau$ spent in each channel $f_j$ is approximately proportional
to the inverse of the average spin-coupling for the $j$-th spin.
(b) The spin packets probability distribution $|c_j|^2$ in the
$\ket{j}$ basis for a selection of times. These are readily
computed from the $x$-axis rotation of $z$ angular momentum states
from which the spread $s_j$ can then be extracted. (c) The
fraction of time $f_j$ (left axis and solid line) and spread $s_j$
(right axis and dashed line) as a function of the spin
$j$.}\label{fig:depolarmodel}
\end{center}
\end{figure}

\subsection{Effective two qubit gate}
\label{twoqubitsec} In this section we consider the EB as a
mediator of an effective two-qubit c-$\sigma^z$ gate. We
systematically compute the average gate fidelity $\av{F}$ for this
operation and the minimum eigenvalue $\varepsilon_{\mathrm{min}}$
of the partial transposition, with respect to the bipartition
$(12)(34)$, of the 4 qubit state $\varrho^{\Lambda}$ isomorphic to
the accumulative noise $\Lambda^{[2]}$. If
$\varepsilon_{\mathrm{min}}<0$ then the noisy operation of the EB
is still capable of entanglement generation. We also use the
behaviour of $\av{F}$ to determine if the accumulative noise is a
product of local noise of the type introduced in section
\ref{noise}.

\subsubsection{Decay noise (low-$T$ limit) - } Earlier in section
\ref{decaynoise} we found that using the EB as a single-qubit
channel with local decay noise results in the accumulative noise
$\Lambda^{[1]}$ being independent of $N$. Thus all chain lengths
were equivalent to a chain with just one spin. When the EB is used
to mediate a c-$\sigma^z$ gate with local decay noise, our numerical
results show that $\av{F}$, for the lengths investigated, is
independent on $N$. However, further investigation reveals that
$\Lambda^{[2]}$ itself does possess a weak length dependence
demonstrating that this result is a consequence of $\av{F}$ being
completely insensitive to these changes. A length dependence of
$\Lambda^{[2]}$ is expected since, in contrast to the single-qubit
channel in section \ref{decaynoise}, the projected master equation
in the subspace $\mathcal{H}_0 \oplus \mathcal{H}_1 \oplus
\mathcal{H}_2$ does not have commuting coherent and dissipative
contributions. Our numerical calculations furthermore indicate that
$\varepsilon_{\mathrm{min}}$ is weakly length dependent but is never
positive over the parameter range and lengths investigated.

We find that the coupling at which the fidelity drops to
$\av{F}<0.99$ is $\kappa_f/J = 4.0\times 10^{-3}$ independent of
$N$. In addition to the average gate fidelity we also compute the
specific gate fidelity $F^{++}$ of the initial state $\ket{++}$
using \eqr{specificfid} and the numerically determined superoperator
$\Lambda^{[2]}$. Using this initial state makes the operation
equivalent to the noisy generation of a two-qubit cluster state. We
find that the coupling at which this fidelity drops to $F^{++}<0.99$
is $\kappa^{++}_f/J = 8.5\times 10^{-3}$, and turns out to be
independent of $N$. Thus, this specific preparation is twice as
resilient to decay noise than the average preparation.

\subsubsection{Dephasing noise -}
For local dephasing noise we find that
$\varepsilon_{\mathrm{min}}$, shown in \fir{fig:dephase_two}(a),
is never positive over the parameter regime considered and
therefore the ability to generate entanglement is retained in the
presence of this noise. In a similar way to the single-qubit
channel fidelity we find that the gate fidelity, plotted in
\fir{fig:dephase_two}(b), rapidly converges with increasing $N$.
To understand the nature of the accumulative noise we attempted to
fit the fidelity to a model where the ideal two-qubit gate $U$ is
implemented and then product noise
$\Lambda^{[1]}_{\mathrm{mod}}\otimes\Lambda^{[1]}_{\mathrm{mod}}$
is applied where $\Lambda^{[1]}_{\mathrm{mod}}$ is a single-spin
superoperator describing noise from the class introduced in
section \ref{noise}. In fact we found that the best fit was
obtained when $\Lambda^{[1]}_{\mathrm{mod}}$ was further
restricted to the case used in section \ref{dephaseone} where
$\gamma=\kappa$ and $\beta = \zeta\kappa$. The validity of these
fits determined from $F_{\Lambda}$ had a peak infidelity of
$1-F_{\Lambda} \approx 0.3$ for strong coupling with $N=2$ and so
the actual noise bears no resemblance to this product model. This
rapidly drops to $1-F_{\Lambda} \approx 5\times 10^{-2}$ for
larger $N$, which indicates that the product noise model becomes
more valid for longer chains. This behavior is sensible since in
long chains the two spin packets do not overlap for the majority
of the evolution time $\tau$ and therefore experience independent
noise during this time. We also find that the coupling at which
the fidelities $\av{F}$ and $F^{++}$ drop to 99\% are $\kappa_f/J
= 2.5\times 10^{-3}$ and $\kappa^{++}_f/J = 6.5\times 10^{-3}$
respectively, for chain lengths up to $N=12$.

\begin{figure}
\begin{center}
\includegraphics[width=11cm]{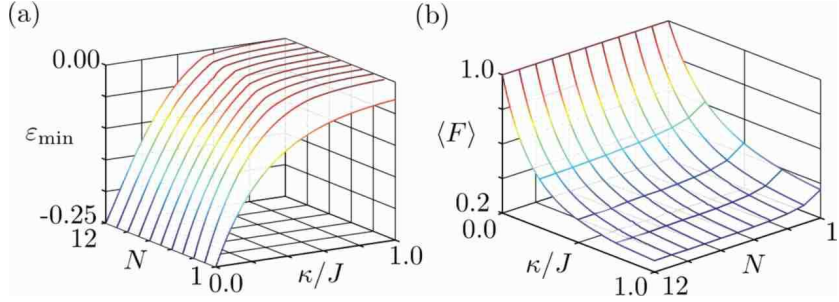}
\caption{For local dephasing noise - (a) The minimum eigenvalue
$\varepsilon_{\mathrm{min}}$ of the partial transposition of the 4
qubit mixed state $\varrho^{\Lambda}$ for the bipartition
$(12)(34)$ and (b) the average gate fidelity $\av{F}$, both as a
function of the chain length $N$ and coupling strength
$\kappa/J$.} \label{fig:dephase_two}
\end{center}
\end{figure}

\subsubsection{Thermal noise (finite-$T$) -}
For local thermal noise we find that $\varepsilon_{\mathrm{min}}$,
shown in \fir{fig:temp_results_two}(a), is only marginally
increased and remains negative over the parameters we considered.
Despite this from our analysis of the single-qubit channel there
is good reason to suspect that for longer chains and higher
temperatures this noise will generate
$\varepsilon_{\mathrm{min}}>0$. Along with $\av{F}$ depicted in
\fir{fig:temp_results_two}(b) $\varepsilon_{\mathrm{min}}$ has a
very weak dependence on $N$ for $T/\Delta<0.2$ similar to that
encountered for a single-qubit channel in section \ref{tempone}.
We again fit the fidelity surface with a product noise model
$\Lambda^{[1]}_{\mathrm{mod}}$ assuming the same single spin noise
as used in section \ref{tempone} where $\alpha(T)$ and $\beta(T)$
remain unchanged from those in \eqr{thermalrates} and $\gamma(T) =
\zeta\alpha(T)$ with a fit parameter $\zeta$. This fitting
produces a linear dependence of $\zeta$ with $N$ as found earlier
for the single qubit channel. In this case the infidelity for the
product noise model is $1-F_{\Lambda}<2.3\times 10^{-2}$ and is
therefore a good approximation to the accumulative noise over the
parameter range investigated. This indicates the noise is very
effective at eliminating correlations which might be built up by
the dynamics of the chain. We also find that the temperatures at
which the fidelities $\av{F}$ and $F^{++}$ drop to 99\% of their
$T=0$ value are $T_f/\Delta = 0.29$ and $T^{++}_f/\Delta = 0.34$
respectively, for chain lengths up to $N=12$.

\begin{figure}
\begin{center}
\includegraphics[width=11cm]{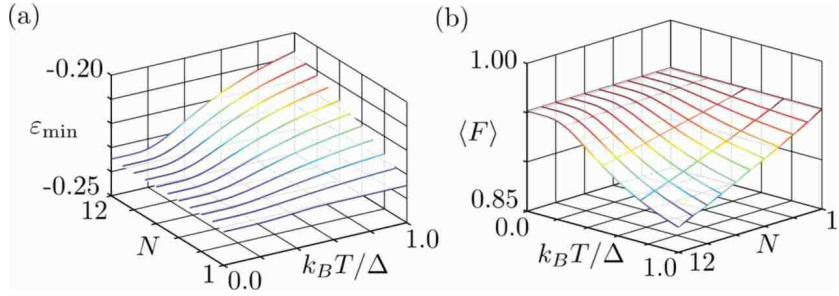}
\caption{For local thermal reservoirs - (a) The minimum eigenvalue
$\varepsilon_{\mathrm{min}}$ of the partial transposition of the 4
qubit mixed state $\varrho^{\Lambda}$ for the bipartition
$(12)(34)$ and (b) the average gate fidelity $\av{F}$, both as a
function of the chain length $N$ and temperature $k_BT/\Delta$
using a coupling $\kappa/J=0.02$} \label{fig:temp_results_two}
\end{center}
\end{figure}

\subsubsection{Depolarizing noise (high-$T$ limit) -}
For local depolarizing noise we observe in
\fir{fig:depolar_two_results}(a) that $\varepsilon_{\mathrm{min}}$
becomes positive for a sizable portion of the parameter range
explored. We can therefore only be certain that entanglement
generation is possible with the EB outside this region of
parameters, i.e. $\kappa/J<0.15$ for chain lengths up to $N=12$.
The average gate fidelity plotted in
\fir{fig:depolar_two_results}(b) shows that $\kappa \ll J$ is
required in order to achieve a reasonable average fidelity. When
fitting a product noise model to $\av{F}$ we obtain an infidelity
$1-F_{\Lambda}<4.5\times 10^{-2}$ which decreases significantly
with larger coupling $\kappa/J$. This is consistent with stronger
local noise destroying correlations between the spin packets and
therefore decorrelating the noise. We also find that the coupling
at which the fidelities $\av{F}$ and $F^{++}$ drop to 99\% are
$\kappa_f/J = 8.0\times 10^{-4}$ and $\kappa^{++}_f/J = 1.4\times
10^{-3}$ respectively, for chain lengths up to $N=12$.

\begin{figure}
\begin{center}
\includegraphics[width=11cm]{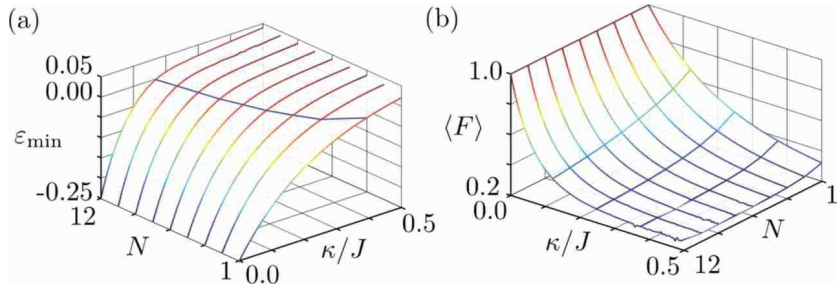}
\caption{For local depolarizing noise - (a) The minimum eigenvalue
$\varepsilon_{\mathrm{min}}$ of the partial transposition of the 4
qubit mixed state $\varrho^{\Lambda}$ for the bipartition
$(12)(34)$ and (b) the average gate fidelity $\av{F}$, both as a
function of the chain length $N$ and coupling strength $\kappa/J$.
The critical coupling $\kappa_c/J$ for each chain length $N$ at
which the entanglement generating capability of the EB is no
longer certain is given by the intersection with the
$\varepsilon_{\mathrm{min}}=0$ plane shown in (a).}
\label{fig:depolar_two_results}
\end{center}
\end{figure}

\section{Implications for graph state generation}
\label{graphgen} In section \ref{graph} a general scheme for
creating arbitrary graph states was outlined which exploits the
multi-qubit circuit implemented by the EB. A complete
characterization of the influence of noise on the full scheme is
beyond the scope of the current work. Here, we instead focus on the
most direct implication of the results presented in section
\ref{twoqubitsec} by investigating graph state generation for five
qubits using the EB to mediate a two-qubit c-$\sigma^z$ only. As an
example we focus on the generation of a linear cluster state, shown
in \fir{fig:graph_gen}(a), and a GHZ state, shown in
\fir{fig:graph_gen}(b), under the influence of local dephasing
noise. These two states were chosen because they have the same
number of edges but very different topologies.

The generation protocol begins by initializing the first five qubits
of a ten qubit register in the state $\ket{+}$. For the previous
calculations in section \ref{twoqubitsec} the input and output
qubits were exclusively the end spins in the EB. For graph state
generation, however, each usage of the EB to establish an edge will
necessarily involve using different spins of the chain as the input
qubits (as well as output qubits). Our numerical calculations
indicate that the superoperator $\Lambda^{[2]}$ does depend weakly
on the input qubit locations. For this reason the order in which the
gates are performed will affect the quality of the overall state. To
illustrate effect this we computed the graph state generation
protocol for two different gate ordering schemes. The first, scheme
($i$), performs the necessary gates in a simple sequential order
according to the qubit labels. The second, scheme ($ii$), performs a
gate between qubits, from the list of edges to be established, which
are closest in the register at a given step (which is often not
unique). These two schemes are shown explicitly for the linear
cluster state in \fir{fig:graph_gen}(c). We computed numerically the
superoperator $\Lambda^{[2]}$ for all input locations required for
these two schemes to generate the linear cluster state and the GHZ
state. By concatenating the appropriate noisy gates in the specified
order the imperfect graph state was obtained.

\begin{figure}
\begin{center}
\includegraphics[width=10cm]{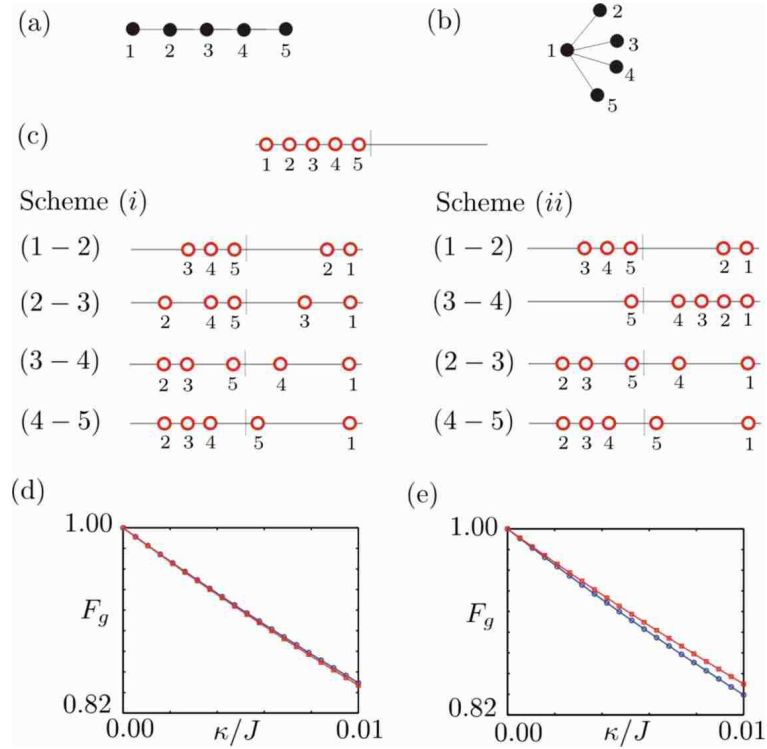}
\caption{We consider the generation of (a) the linear cluster
state and (b) the GHZ state for five qubits. The qubits
representing the graph vertices are initialized in the first half
of the register as shown in (c). Two schemes for the sequence in
which the EB is used to mediate the necessary two-qubit gates are
investigated. Scheme ($i$) proceeds with the gates in a sequential
order, while scheme ($ii$) proceeds by performing gates between
qubits which are closest in the register at a given step. For the
linear cluster state the two schemes are explicitly shown in (c).
In (d) the graph state fidelity $F_g$ for the linear cluster state
(`$\circ$') and GHZ state (`$\Box$') constructed with scheme ($i$)
is plotted, and in (e) the same is shown for scheme ($ii$).}
\label{fig:graph_gen}
\end{center}
\end{figure}

For scheme ($i$), shown in \fir{fig:graph_gen}(d), we find that the
linear cluster state fidelity $F_g$ is slightly above that for the
GHZ state with them having a fidelity greater than 99\% for
$\kappa/J < 5.7 \times 10^{-4}$ and $\kappa/J < 6.0 \times 10^{-4}$,
respectively. For scheme ($ii$), shown in \fir{fig:graph_gen}(e),
the GHZ state fidelity is virtually unchanged from that of scheme
($i$), whereas the linear cluster state fidelity has dropped below
that of the GHZ state giving a slightly reduced $\kappa/J < 6.1
\times 10^{-4}$ for a fidelity greater than 99\% to be attained. The
average gate fidelity for a $N=10$ chain, computed in section
\ref{twoqubitsec} using the end spins as the input qubits, drops to
99\% when $\kappa/J < 2.6 \times 10^{-3}$ for local dephasing noise.
Thus the noise tolerance for the generation of these graph states is
reduced by a factor of approximately $4.5$ in comparison.
Additionally, the results appear to indicate that only a weak change
in the fidelity is observed with the graph topology and input qubit
locations for a $N=10$ chain. Further work is need to confirm if
this insensitivity is maintained for more varied topologies over
larger numbers of vertices and longer chains.

These preliminary results give a clear indication of useful
directions for future work. This includes studying the dependence of
the EB on the input qubit locations with different noise models as
well as chain lengths and determining wether there is any generic
behaviour. For the full graph state generation scheme of section
\ref{graph} the quality of the multi-qubit circuits implemented by
the EB with noise needs to be studied along with their likely
dependence on the input qubit locations. This may well reveal a
trade-off between using the EB less times with more qubits or more
times with fewer qubits. With this information and for a given graph
topology, the graph state generation scheme under the influence of
noise could be optimized, in terms of the sequence and type of gates
implemented.

\section{Conclusion}
\label{conclusion} In summary we have investigated the influence of
local noise on the mirror inverting EB spin chain. For the case
where the EB is utilized as a single qubit channel we have found
that the accumulative noise of the EB for local decay noise is
independent of its length $N$ and explained this unexpected
behaviour. Additionally, we have found that neither local decay nor
dephasing noise cause the EB to become entanglement breaking, in
contrast to local thermal and depolarizing noise. For the latter two
cases we have determined the critical length $N_c$ and critical
coupling $\kappa_c/J$ at which entanglement breaking occurs,
respectively. The local depolarizing noise $\kappa_c/J$ is found to
exhibit a power-law dependence on $N$ which is explained by the
competition between the speed and spreading of the spin packet in
the EB.

For the case where the EB is used to mediate an entangling
c-$\sigma^z$ gate we find that the entanglement generating
capability of the EB is never lost in presence of local decay or
dephasing noise. For local dephasing noise the resulting operation
of the EB becomes progressively closer to product noise with
increasing $N$. Both local thermal and depolarizing noise are well
approximated by product noise models due to the severity at which
they decohere the spin chain. For local depolarizing noise our
results indicate that the entanglement generating capability of
the EB can only be guaranteed for couplings $\kappa/J<0.15$ for
chain lengths up to $N=12$.

We have also performed a preliminary analysis of the graph state
generation scheme which uses the EB, with local dephasing noise, to
individually implement the necessary c-$\sigma^z$ gates for a five
qubit graph state. As an example we focussed on the linear cluster
state and GHZ state. As expected we found that the concatenation of
these operations reduces the tolerance to noise. In this case the
reduction was by a factor of approximately $4.5$ for both states
compared to the average fidelity of a single gate. More work is
needed to determine how this behaviour scales for larger graph
states constructed with longer EB chains. Finally, we note that
while our results indicate that there are tight constraints on the
levels of tolerable noise for graph state generation this can be
weakened within the framework of one-way quantum computing due to
the separation between the preparation and consumption of
entanglement~\cite{Tame-PRA-2005}. In principle this enables the
resulting noisy graph states to be purified prior to their
use~\cite{Dur-PRL-2003,Aschauer-2005}.

\ack SRC thanks Richard Walters for helpful discussions. This work
is supported by the UK EPSRC through QIP IRC (GR/S82176/01) and
project EP/C51933/1, the Berrow Scholarship (MB), and the Keble
Association (AK).

\appendix

\section{Limits of the local noise model}
\label{limits}

In section \ref{noise} the general form for the Kraus operators was
given in \eqr{kraus} for the noise model considered here. Since the
expressions for the general matrix elements $\Upsilon_i$ of the
diagonal operators are rather lengthy we have postponed their
introduction to this appendix. To begin with the following
quantities are defined~\cite{Hein-PRA-2005}
\begin{eqnarray}
\lambda_0(t) &=& \frac{1}{4}[1+2e^{-\frac{1}{2}(\alpha + \beta +
2\gamma)t} + e^{-(\alpha + \beta)t}], \nonumber \\
\lambda_1(t) &=& \lambda_2(t) = \frac{1}{4}[1 - e^{-(\alpha + \beta)t}], \nonumber \\
\lambda_3(t) &=& \frac{1}{4}[1-2e^{-\frac{1}{2}(\alpha + \beta +
2\gamma)t} + e^{-(\alpha + \beta)t}], \nonumber \\
\mu(t) &=& \frac{\av{\sigma_z}_s}{4}[1 - e^{-(\alpha + \beta)t}],
\nonumber
\end{eqnarray}
using the definitions from section \ref{noise}. We then find that
for $\mu \neq 0$
\begin{eqnarray}
\Upsilon_1 &=& \frac{1}{2} (x-2 \mu-\lambda_0+\lambda_3)
\sqrt{\frac{\lambda_0
  +\lambda_3-x}{4 \mu ^2-(\lambda_0-\lambda_3)
  (x-\lambda_0+\lambda_3)}} \nonumber \\
\Upsilon_2 &=& \frac{1}{2} (2 \mu +x-\lambda_0+\lambda_3)
\sqrt{\frac{\lambda_0+\lambda_3-x}
 {4 \mu ^2-(\lambda_0-\lambda_3) (x-\lambda_0+\lambda_3)}} \nonumber \\
\Upsilon_3 &=& \frac{1}{2} (2 \mu +x+\lambda_0-\lambda_3)
\sqrt{\frac{x+\lambda_0+\lambda_3}
 {4 \mu ^2+(\lambda_0-\lambda_3) (x+\lambda_0-\lambda_3)}} \nonumber \\
\Upsilon_4 &=& \frac{1}{2} (x-2 \mu+\lambda_0-\lambda_3)
\sqrt{\frac{x+\lambda_0+\lambda_3}
   {4 \mu ^2+(\lambda_0-\lambda_3) (x+\lambda_0-\lambda_3)}}
   \nonumber
\end{eqnarray}
making one further definition $x = \sqrt{4 \mu^2 + (\lambda_0 -
\lambda_3)^2}$. As expected the Kraus operators in \eqr{kraus}
simplify considerably for a number of important special cases. In
particular for $\mu=0$ the Kraus operators become $E_1 =
\sqrt{\lambda_0}\,\mathbbm{1}$, $E_2 =
\sqrt{\lambda_1}\,\sigma^x$, $E_3 = \sqrt{\lambda_2}\,\sigma^y$,
$E_4 = \sqrt{\lambda_3}\,\sigma^z$. For $\beta = \kappa$ and
$\alpha=\gamma=0$ they reduce to
\begin{eqnarray}
E_{1} &=& \left(\begin{array}{cc}
1 & 0  \\
0 & \sqrt{p_1}
\end{array}\right),\quad E_{2} = \left(\begin{array}{cc}
0 & \sqrt{1-p_1}  \\
0 & 0
\end{array}\right), \nonumber
\end{eqnarray}
with $p_1 = e^{-\kappa t}$ describing pure decay noise; for
$\gamma = \kappa$ and $\alpha=\beta=0$ they reduce to
$E_1=\sqrt{p_2}\,\mathbbm{1}$, $E_2 = \sqrt{1-p_2}\,\sigma^z$ with
$p_2=\sqrt{\half(1+e^{-\kappa t})}$ describing pure dephasing;
while for $\alpha = \beta = \gamma = \half\kappa$ they become $E_1
= \sqrt{1-3p_3}\,\mathbbm{1}$, $E_2 = \sqrt{p_3}\,\sigma^x$, $E_3
= \sqrt{p_3}\,\sigma^y$, $E_4 = \sqrt{p_3}\,\sigma^z$ with $p_3 =
\frac{1}{4}(1-e^{-\kappa t})$ describing depolarizing
noise~\cite{Nielsen-2000}. For finite-$T$ noise with $\gamma=0$
and $\alpha+\beta=\kappa$ the Kraus operators take the form
\begin{eqnarray}
E_{1} &=& \sqrt{P_{\uparrow}} \left(\begin{array}{cc}
1 & 0  \\
0 & \sqrt{e^{-\kappa t}}
\end{array}\right),\quad E_{2} = \sqrt{P_{\uparrow}}\left(\begin{array}{cc}
0 & \sqrt{1-e^{-\kappa t}}  \\
0 & 0
\end{array}\right), \nonumber \\
E_{3} &=& \sqrt{P_{\downarrow}} \left(\begin{array}{cc}
\sqrt{e^{-\kappa t}} & 0 \\
0 & 1 \end{array}\right), \quad E_{4} =
\sqrt{P_{\downarrow}}\left(\begin{array}{cc}
0 & 0  \\
\sqrt{1-e^{-\kappa t}} & 0 \end{array}\right). \nonumber
\end{eqnarray}

For $\mu \neq 0$ the average fidelity for a single-qubit channel
experiencing this class of noise can be computed from \eqr{fidelity}
as
\begin{eqnarray}
\av{F} &=& \frac{1}{6} \left( 2 +
  \frac{(\lambda_0 - \lambda_3 +x)^2 (\lambda_0 + \lambda_3 + x)}
  {4 \mu^2 +(\lambda_0 - \lambda_3)(\lambda_0 - \lambda_3 +x)}
  + \frac{(\lambda_0 - \lambda_3 -x)^2 |\lambda_0 + \lambda_3 - x|}
  {|4 \mu^2 - (\lambda_0-\lambda_3)(\lambda_3-\lambda_0+x)|}
  \right). \nonumber
\end{eqnarray}
For finite-$T$ noise with $\gamma=0$ this equation reduces to
\begin{eqnarray}
\av{F_{T}} &=& \frac{1}{2} + \frac{1}{3}e^{-\half\kappa t} +
\frac{1}{6}e^{-\kappa t}, \nonumber
\end{eqnarray}
and has a double exponential decay independent of $T$ making it
equally applicable to the $T=0$ decay channel. For the special
case where $\mu = 0$ the expression for $\av{F}$ dramatically
simplifies to
\begin{eqnarray}
\av{F_{\mu=0}} &=& \frac{1}{6} \left(4 \lambda_0 + 2\right),
\nonumber
\end{eqnarray}
from which the well known fidelities~\cite{Nielsen-2000} for
dephasing and depolarizing channels can easily be evaluated as
\begin{eqnarray}
\av{F_{\mathrm{dephase}}} &=& \frac{2}{3} + \frac{1}{3}e^{-\kappa
t}, \nonumber
\\
\av{F_{\mathrm{depolar}}} &=& \frac{1}{2} + \frac{1}{2}e^{-\kappa
t}, \nonumber
\end{eqnarray}
which exponentially decay to $\frac{2}{3}$ and $\half$,
respectively.

\section{Positive partial transposition criterion}
\label{ppt}

Detecting the presence of entanglement in a general bipartite mixed
state can be achieved to an extent through the use of the positive
partial transposition (PPT)
criterion~\cite{Peres-PRL-1996,Horodecki-PLA-1996}. For an operator
$O$ acting on the Hilbert space $\mathcal{H} =
\mathbbm{C}^{\mathbbm{d}_a}\otimes\mathbbm{C}^{\mathbbm{d}_b}$ of
two systems $a$ and $b$ the partial transpose with respect to system
$a$ is defined as
\begin{eqnarray}
O^{T_a} = \sum_{ij=1}^{\mathbbm{d}_a}\sum_{mn=1}^{\mathbbm{d}_b}
\bracket{i,m}{O}{j,n}\outprod{j,m}{i,n}, \nonumber
\end{eqnarray}
in terms of some basis $\{\ket{i,m}~|~ i=1,\cdots,\mathbbm{d}_a,
m=1,\cdots,\mathbbm{d}_b\}$ of $\mathcal{H}$. Although this
definition is basis-dependent the spectrum of $O^{T_a}$ is not. Note
also that while the transposition of the full system $ab$ preserves
the positivity of the full density matrix $(\rho^{T_a})^{T_b} =
\rho^T \geq 0$, in general the transposition with respect to any
subsystem does not, and so the partial transpose is not a completely
positive operation.

A state $\rho$ of the system $ab$ is separable if and only if it
can be expressed as a convex combination of product states
\begin{eqnarray}
\rho &=& \sum_{i=1}^{\chi}p_i\rho^a_i\otimes\rho^b_i, \nonumber
\end{eqnarray}
with $p_i \geq 0$ and $\sum_{i=1}^\chi p_i = 1$. The PPT criterion
then states that $\rho^{T_a} \geq 0$ is a necessary condition for
separability of any $d_a \times d_b$ systems. Importantly for $2
\times 2$ and $2 \times 3$ systems the PPT criterion is necessary
and sufficient for
separability~\cite{Peres-PRL-1996,Horodecki-PLA-1996}.

\section{Jamiolkowski isomorphism}
\label{Jamiolkowski}

In section \ref{character} extensive use of the Jamiolkowski
isomorphism~\cite{Jamiolkowski-RMP-1972,Hein-PRA-2005} is made. This
isomorphism establishes an equivalence between quantum states and
superoperators. To begin suppose we have a type of subsystem $S$
with Hilbert space $\mathcal{H}$ of dimension $\mathbbm{d}$ and
spanned by basis states $\{\ket{i}~|~ i=0,\cdots,\mathbbm{d}-1\}$.
For such a subsystem any density operator $\rho$ can be expanded in
the operator basis $\{\outprod{i}{j}~|~
i,j=0,\cdots,\mathbbm{d}-1\}$ with its corresponding matrix elements
contained in a $\mathbbm{d}^2$-dimensional vector $\rho_{ij}$.
Superoperators are defined as linear, trace-preserving, completely
positive maps of density operators to density operators.
Consequently, a superoperator $\Lambda^{[1]}$ acting on a subsystem
$S$ is completely described by a $\mathbbm{d}^2 \times
\mathbbm{d}^2$ super-matrix with elements in the operator basis
$\Lambda^{[1]}_{ij,kl}$ as
\begin{eqnarray}
\Lambda^{[1]}\{\outprod{i}{j}\} &=& \sum_{kl=0}^{\mathbbm{d}-1}
\Lambda^{[1]}_{ij,kl}\outprod{k}{l}. \nonumber
\end{eqnarray}
This information can be mapped to a quantum state by using two
copies $a$ and $b$ of the subsystem $S$ initially prepared in the
maximally entangled state $\ket{\Psi} =
\frac{1}{\sqrt{\mathbbm{d}}}\sum_{i=0}^{\mathbbm{d}-1}\ket{i}\otimes\ket{i}$
and applying the superoperator $\Lambda^{[1]}$ to $b$ as
\begin{eqnarray}
(\mathbbm{1}\otimes\Lambda^{[1]})\{\outprod{\Psi}{\Psi}\} &=&
\frac{1}{\mathbbm{d}}\sum_{ij=0}^{\mathbbm{d}-1}
\outprod{i}{j}\otimes\Lambda^{[1]}\{\outprod{i}{j}\} =
\rho^{\Lambda}, \nonumber
\end{eqnarray}
as shown in \fir{isomorphism}(a). The resulting density operator
$\rho^{\Lambda}$ for the two $S$ subsystems then completely
describes $\Lambda^{[1]}$ by containing all of its operator matrix
elements $\Lambda^{[1]}\{\outprod{i}{j}\}$. This can then be used
to compute $\Lambda^{[1]}\{\rho^{\mathrm{in}}\} =
\rho^{\mathrm{out}}$ on any state $\rho^{\mathrm{out}}$ via
\begin{eqnarray}
\Lambda^{[1]}\{\rho^{\mathrm{in}}\} &=&
\mathbbm{d}\sum_{ij=0}^{\mathbbm{d}-1}
\left(\,\sum_{kl=0}^{\mathbbm{d}-1} \rho^{\mathrm{in}}_{kl}
\,\rho^{\Lambda}_{ki,lj}\right)\outprod{i}{j} =
\rho^{\mathrm{out}},\nonumber
\end{eqnarray}
thereby giving the inverse isomorphism.

For a superoperator $\Lambda^{[2]}$ acting on two $S$ subsystems
the Jamiolkowski isomorphism proceeds in an analogous way by
applying $\Lambda^{[2]}$ on one half of two  maximally entangled
pairs of subsystems with the setup depicted in
\fir{isomorphism}(b) as
\begin{eqnarray}
(\mathbbm{1}_a\otimes\Lambda^{[2]}_{bc}\otimes\mathbbm{1}_d)\{\outprod{\Psi}{\Psi}_{ab}\otimes\outprod{\Psi}{\Psi}_{cd}\}
&=& \varrho^{\Lambda} \nonumber.
\end{eqnarray}
The resulting density matrix $\varrho^{\Lambda}$ for the four $S$
subsystems is then
\begin{eqnarray}
\varrho^{\Lambda} &=&
\frac{1}{\mathbbm{d}^2}\sum_{ijkl=0}^{\mathbbm{d}-1}
\outprod{i}{j}\otimes\Lambda^{[2]}\{\outprod{ik}{jl}\}\otimes\outprod{k}{l}
\nonumber,
\end{eqnarray}
and again completely describes $\Lambda^{[2]}$ through its matrix
elements in the operator basis of two subsystems.

\section*{References}

\bibliographystyle{iopart-num}
\bibliography{SpinChain}

\providecommand{\newblock}{}
\begin{thebibliography}{10}
\expandafter\ifx\csname url\endcsname\relax
  \def\url#1{{\tt #1}}\fi
\expandafter\ifx\csname urlprefix\endcsname\relax\def\urlprefix{URL }\fi
\providecommand{\eprint}[2][]{\url{#2}}

\bibitem{Jozsa-PRSA-2003}
Jozsa R and Linden N 2003 {\em Proc. R. Soc. A\/} {\bf 459} 2011

\bibitem{Vidal-PRL-2003}
Vidal G 2003 {\em Phys. Rev. Lett.\/} {\bf 91} 147902

\bibitem{Raussendorf-PRL-2001}
Raussendorf R and Briegel H~J 2001 {\em Phys. Rev. Lett.\/} {\bf 86} 5188

\bibitem{Raussendorf-PRA-2003}
Raussendorf R, Browne D~E and Briegel H~J 2003 {\em Phys. Rev. A\/} {\bf 68}
  022312

\bibitem{Hein-PRA-2004}
Hein M, Eisert J and Briegel H~J 2004 {\em Phys. Rev. A\/} {\bf 69} 062311

\bibitem{Hein-2006}
Hein M, D{\"u}r W, Eisert J, Raussendorf R, {Van den Nest} M and Briegel H~J
  2006  (\textit{Preprint} \eprint{quant-ph/0602096})

\bibitem{Hein-PRA-2005}
Hein M, D{\"u}r W and Briegel H~J 2005 {\em Phys. Rev. A\/} {\bf 71} 032350

\bibitem{Briegel-PRL-2001}
Briegel H~J and Raussendorf R 2001 {\em Phys. Rev. Lett.\/} {\bf 86} 910

\bibitem{Gottesman-1997}
Gottesman D 1997  (\textit{Preprint} \eprint{quant-ph/9705052})

\bibitem{Browne-PRL-2005}
Browne D~E and Rudolph T 2005 {\em Phys. Rev. Lett.\/} {\bf 95} 010501

\bibitem{Walther-NAT-2005}
Walther P, Resch K~J, Rudolph T, Schenck E, Weinfurter H, Vedral V, Aspelmeyer
  M and Zeilinger A 2005 {\em Nature\/} {\bf 434} 169

\bibitem{Barrett-PRA-2005}
Barrett S~D and Kok P 2005 {\em Phys. Rev. A\/} {\bf 71} 060310(R)

\bibitem{Bartlett-PRA-2006}
Bartlett S~D and Rudolph T 2006 {\em Phys. Rev. A\/} {\bf 74} 040302(R)

\bibitem{Christandl-PRL-2004}
Christandl M, Datta N, Ekert A and Landahl A~J 2004 {\em Phys. Rev. Lett.\/}
  {\bf 92} 187902

\bibitem{Christandl-PRA-2005}
Christandl M, Datta N, Dorlas T~C, Ekert A, Kay A and Landahl A~J 2005 {\em
  Phys. Rev. A\/} {\bf 71} 032312

\bibitem{Cook-PRA-1979}
Cook R~J and Shore B~W 1979 {\em Phys. Rev. A\/} {\bf 20} 539

\bibitem{Yung-PRA-2005}
Yung M~H and Bose S 2005 {\em Phys. Rev. A\/} {\bf 71} 032310

\bibitem{Karbach-PRA-2005}
Karbach P and Stolze J 2005 {\em Phys. Rev. A\/} {\bf 72} 030301(R)

\bibitem{Yung-PRA-2006}
Yung M~H 2006 {\em Phys. Rev. A\/} {\bf 74} 030303

\bibitem{Kay-PRL-2007}
Kay A 2007 {\em Phys. Rev. Lett.\/} {\bf 98} 010501

\bibitem{Clark-NJP-2005}
Clark S~R, {Moura Alves} C and Jaksch D 2005 {\em New Journal of Physics\/}
  {\bf 7} 124

\bibitem{Breuer-2002}
Breuer H~P and Petruccione F 2002 {\em The theory of open quantum systems\/}
  1st ed (Oxford: Oxford University Press)

\bibitem{DeChiara-PRA-2005}
Chiara G~D, Rossini D, Montangero S and Fazio R 2005 {\em Phys. Rev. A\/} {\bf
  72} 012323

\bibitem{Cai-PRA-2006}
Cai J~M, Zhou Z~W and Guo G~C 2006 {\em Phys. Rev. A\/} {\bf 74} 022328

\bibitem{Burgarth-PRA-2006}
Burgarth D and Bose S 2006 {\em Phys. Rev. A\/} {\bf 73} 062321

\bibitem{Zhou-2006}
Zhou L, Lu J, Shi T and Sun C~P 2006  (\textit{Preprint}
  \eprint{quant-ph/0608135})

\bibitem{Briegel-PRA-1993}
Briegel H~J and Englert B~G 1993 {\em Phys. Rev. A\/} {\bf 47} 3311

\bibitem{Sachdev-1999}
Sachdev S 1999 {\em Quantum Phase Transitions\/} 1st ed (Cambridge: Cambridge
  University Press)

\bibitem{Plenio-NJP-2004}
Plenio M~B, Hartley J and Eisert J 2004 {\em New J. Phys.\/} {\bf 6} 36

\bibitem{Chaichian-1998}
Chaichian M and Hagedorn R 1998 {\em Symmetries in Quantum Mechanics\/} 1st ed
  (London: Institute of Physics)

\bibitem{Nielsen-2000}
Nielsen M and Chuang I~L 2000 {\em Quantum Computation and Quantum
  Information\/} 1st ed (Cambridge: Cambridge University Press)

\bibitem{Dankert-2005}
Dankert C 2005  (\textit{Preprint} \eprint{quant-ph/0512217})

\bibitem{Pedersen-2007}
Pedersen L~H, M{\o}lmer K and M{\o}ller N~M 2007  (\textit{Preprint}
  \eprint{quant-ph/0701138})

\bibitem{Peres-PRL-1996}
Peres A 1996 {\em Phys. Rev. Lett.\/} {\bf 77} 1413

\bibitem{Horodecki-PLA-1996}
Horodecki M, Horodecki P and Horodecki R 1996 {\em Phys. Lett. A\/} {\bf 223} 1

\bibitem{Jamiolkowski-RMP-1972}
Jamiolkowski A 1972 {\em Rep. Math. Phys.\/} {\bf 3} 275

\bibitem{Zwolak-PRL-2004}
Zwolak M and Vidal G 2004 {\em Phys. Rev. Lett.\/} {\bf 93} 207205

\bibitem{Vidal-PRL-2004}
Vidal G 2004 {\em Phys. Rev. Lett.\/} {\bf 93} 040502

\bibitem{Tame-PRA-2005}
Tame M~S, Paternostro M, Kim M~S and Vedral V 2005 {\em Phys. Rev. A\/} {\bf
  72} 012319

\bibitem{Dur-PRL-2003}
D\"ur W, Aschauer H and Briegel H~J 2003 {\em Phys. Rev. Lett.\/} {\bf 91}
  107903

\bibitem{Aschauer-2005}
Aschauer H, D\"ur W and Briegel H~J 2005 {\em Phys. Rev. A\/} {\bf 71} 012319

\end{thebibliography}

\end{document}